\documentclass[%
 reprint,
 amsmath,amssymb,
 aps,
]{revtex4-2}

\usepackage{graphicx}% Include figure files
\graphicspath{{./Figs/}}
\usepackage{dcolumn}% Align table columns on decimal point
\usepackage{bm}% bold math
\begin{document}

\preprint{APS/123-QED}

\title{Folding instabilities in non-Newtonian viscous sheets: \\ shear thinning and shear thickening effects}% Force line breaks with \\
%\thanks{A footnote to the article title}%

\author{Anselmo Pereira}
\thanks{anselmo.soeiro\_pereira@mines-paristech.fr}
\affiliation{PSL Research University, MINES ParisTech, Centre for material forming (CEMEF), CNRS UMR 7635, CS 10207 rue Claude Daunesse, 06904 Sophia-Antipolis Cedex, France}

\author{Nicolas Valade}
%\thanks{nicolas.valade\_pereira@mines-paristech.fr}
\affiliation{PSL Research University, MINES ParisTech, Centre for material forming (CEMEF), CNRS UMR 7635, CS 10207 rue Claude Daunesse, 06904 Sophia-Antipolis Cedex, France}%

\author{Elie Hachem}
\affiliation{PSL Research University, MINES ParisTech, Centre for material forming (CEMEF), CNRS UMR 7635, CS 10207 rue Claude Daunesse, 06904 Sophia-Antipolis Cedex, France}%

\author{Rudy Valette}
\affiliation{PSL Research University, MINES ParisTech, Centre for material forming (CEMEF), CNRS UMR 7635, CS 10207 rue Claude Daunesse, 06904 Sophia-Antipolis Cedex, France}%

\date{\today}% It is always \today, today,
             %  but any date may be explicitly specified

%%%%%%%%%%%%%%%%%%%%%%%%%%%%%%%%%%%%%%%%%%%%%%%%%%%%%%%%%%%%%%%%%%%%%%%%%%%%%%
%%%%%%%%%%%%%%%%%%%%%%%%%%%%%%%%%%%%%%%%%%%%%%%%%%%%%%%%%%%%%%%%%%%%%%%%%%%%%%
\begin{abstract}
In this work, we extend the analyses devoted to Newtonian viscous fluids previously reported by Ribe [Physical Review E \textbf{68}, 036305 (2003)], by investigating shear thickening (dilatant) and shear thinning (pseudoplastic) effects on the development of folding instabilities in non-Newtonian viscous sheets of which viscosity is given by a power-law constitutive equation. Such instabilities are trigged by compression stresses acting on viscous sheets that leave a channel at a very small initial velocity, fall, and then hit a solid surface or a fluid substrate. Our study is conducted through a mixed approach combining direct numerical simulations, energy budget analyses, scaling laws, and experiments. The numerical results are based on an adaptive variational multi-scale method for multiphase flows, while Carpobol gel sheets are considered for the conducted experiments. Two folding regimes are observed: (1) the viscous regime; and (2) the gravitational one. Interestingly, only the latter is affected by shear thinning/thickening manifestations within the material. In short, when gravity is balanced by viscous forces along the non-Newtonian viscous sheet, both the folding amplitude and the folding frequency are given by a power-law function of the sheet slenderness, the Galileo number (the ratio of the gravitational stress to the viscous one), and the flow behaviour index. Highly shear thickening materials develop large amplitude (and low frequency) instabilities, which, in contrast, tend to be suppressed by shear thinning effects, and eventually cease. Lastly, non-Newtonian effects on folding onset/cessation are also carefully explored. As a result, non-Newtonian folding onset and cessation criteria are presented.    
\end{abstract}
%%%%%%%%%%%%%%%%%%%%%%%%%%%%%%%%%%%%%%%%%%%%%%%%%%%%%%%%%%%%%%%%%%%%%%%%%%%%%%
%%%%%%%%%%%%%%%%%%%%%%%%%%%%%%%%%%%%%%%%%%%%%%%%%%%%%%%%%%%%%%%%%%%%%%%%%%%%%%

\maketitle

%%%%%%%%%%%%%%%%%%%%%%%%%%%%%%%%%%%%%%%%%%%%%%%%%%%%%%%%%%%%%%%%%%%%%%%%%%%%%%
%%%%%%%%%%%%%%%%%%%%%%%%%%%%%%%%%%%%%%%%%%%%%%%%%%%%%%%%%%%%%%%%%%%%%%%%%%%%%%
%--------------------------------------------------------------------------------------------------------------------------------------------------------------------------------------%--------------------------------------------------------------------------------------------------------------------------------------------------------------------------------------
\section{Introduction} \label{INTRO}
%--------------------------------------------------------------------------------------------------------------------------------------------------------------------------------------%--------------------------------------------------------------------------------------------------------------------------------------------------------------------------------------
The buckling instability represents one of most fascinating phenomena in fluid mechanics, being typically observed when free liquid asymmetric filaments, sheets and/or jets are exposed to compression stresses \citep{Barnes-58}. Since the energy related to the folding/coiling deformation becomes smaller than the cost of compression \citep{Taylor-69, Cruickshank-88, Yarin-96, Mahadevan-98, Mahadevan-00}, slender viscous fluid filaments tend to buckle, beyond a critical axial load. For small Newtonian fluid filaments compressed at a very small Reynolds number (negligible inertial force), for instance, the folding deformation emerges from a competition between geometrical, surface tension and viscous effects \citep{Merrer-12}. In addition, Newtonian viscous jet columns can stretch, bend and twist when hitting a surface or a substrate at higher velocities, following the balance between viscous, gravitational, capillary and inertial forces \citep{Ribe-03, Ribe-06, Ribe-12, Tian-20}. Such instabilities are observed in a variety of contexts, which includes glass plate fabrication \citep{Pilkington-69}, polymer processing \citep{Pearson-85}, food processing, high-resolution extrusion-based printing \citep{Tian-20} and folding of geological structures \citep{Griffiths-88, Johnson-94}.

\begin{figure*}%[h!]
\centering
\includegraphics[angle=0, scale=0.205]{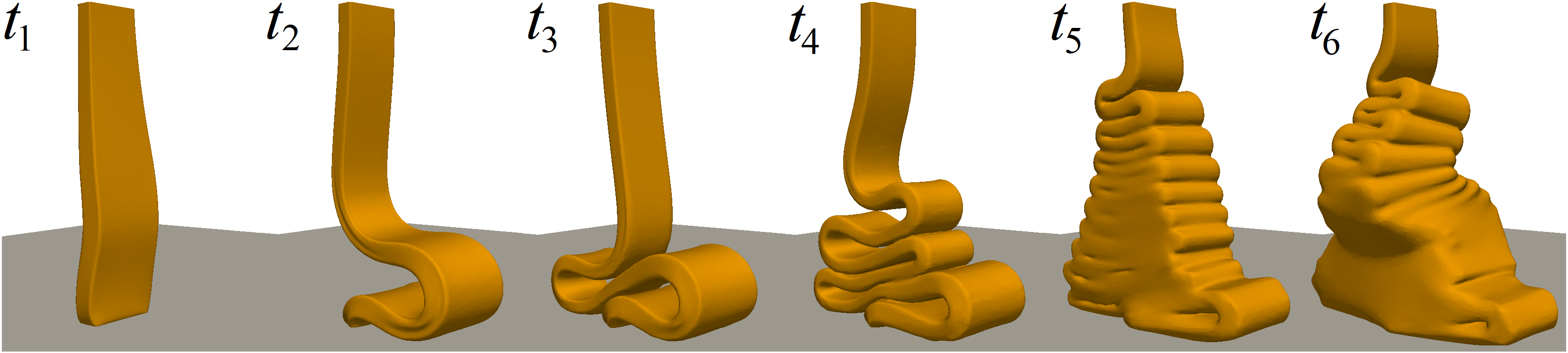}
\vspace*{-0.7cm}
\caption{Three dimensional numerical simulation: time-evolution of the periodic folding of a viscous sheet ($\eta=$ 10 Pa s; $\rho=$ 1000 kg/m$^3$; similar to honey). The height fall is 2.5 cm, and the dimensions of the extrusion slot are $a_0 = 0.14$ cm and $b_0 = 0.5$ cm. Snapshots of six different instant are displayed ($t_1 ~ < ~ t_2 ~ < ~ t_3 ~ < ~ t_4 ~ < ~ t_5 ~ < ~ t_6$). The superposition of several folds, consecutively formed as a result of the fluid filament compression, originates a fluid column ($t_1$-$t_5$), which later on eventually collapses ($t_6$). A movie showing this process is available (movie-1).}
\label{fig-1}
\end{figure*}

In industry, the buckling instability represents a major source of irregularities for container-filling processes \citep{Rasschaert-18}. Typically, as illustrated in Fig. \ref{fig-1}, during these processes, the superposition of several folds/coils, consecutively formed as a result of the fluid filament compression, originates a fluid column completely surrounded by air ($t_1$-$t_5$). Later on, this column eventually collapses (instant $t_6$ in Fig. \ref{fig-1}), entraining a significant amount of air towards the fluid substrate and compromising the quality of the final product. Hence, understanding and controlling the folding/coiling instabilities when dealing with this kind of processes is crucial. Despite some recent works concerning these instabilities in Newtonian contexts \citep{Merrer-12, Habibi-14, Ribe-17} many aspects of the problem remain unclear, such as the effects of non-Newtonian signatures (pseudoplasticity, dilatancy, thixotropy, yield stress etc.) on them \citep{Tome-19, Pereira-19b}.

In the present work, we study the folding process of shear thinning, Newtonian and shear thickening fluid sheets, of which viscosity is given by a power-law constitutive equation \citep{Ostwald-25, Bird-87}. Hence, this study extends earlier works on Newtonian periodic folding \citep{Skorobogatiy-00, Ribe-03} by identifying shear thinning/thickening effects on buckling instabilities. The folding onset, development in time, and cessation are carefully analysed both numerically and experimentally. The performed three-dimensional numerical simulations are based on an adaptive variational multi-scale method for two materials (air and non-Newtonian fluid) combined with a level-set function to provide a precise evolution of the phase interfaces. Numerical results are compared to experimental ones obtained by considering a 5.2 Pa s honey (Newtonian) and four different Carbopol gel sheets (non-Newtonian). Both the folding frequency and amplitude are explored in the light of energy budget analyses and scaling laws. According to these analyses, two folding regimes are observed: (1) the viscous regime; and (2) the gravitational one. Interestingly, only the latter is affected by shear thinning/thickening manifestations within the material. More specifically, in the gravity-driven folding regime, both the folding amplitude and the folding frequency are given by a power-law function of the sheet slenderness, the Galileo number (the ratio of the gravity stress to the viscous one; also called buoyancy number), and the flow behaviour index. Shear thickening materials tend to develop large amplitude (and low frequency) folding instabilities, which, in contrast, tend to be suppressed by shear thinning effects and eventually vanish. Lastly, non-Newtonian effects on both folding onset and cessation are also carefully highlighted. A good agreement between theoretical predictions, three dimensional numerical simulations, and experiments are observed.  

The organisation of the paper is as follows. The descriptions of the physical formulation, the used numerical method and the experimental procedure are given in Section \ref{PFNM}. Our results are discussed in Section \ref{RD}, where two main topics are explored: the folding regimes (Subsection \ref{FR}) and the folding suppression (Subsection \ref{FC}) in power-law sheets. Finally, conclusions are drawn in the closing section.

%--------------------------------------------------------------------------------------------------------------------------------------------------------------------------------------%--------------------------------------------------------------------------------------------------------------------------------------------------------------------------------------
\section{Physical Formulation, Numerical Method and Experimental Procedure} \label{PFNM}
\begin{figure*}%[h!]
\centering
\includegraphics[angle=0, scale=0.307]{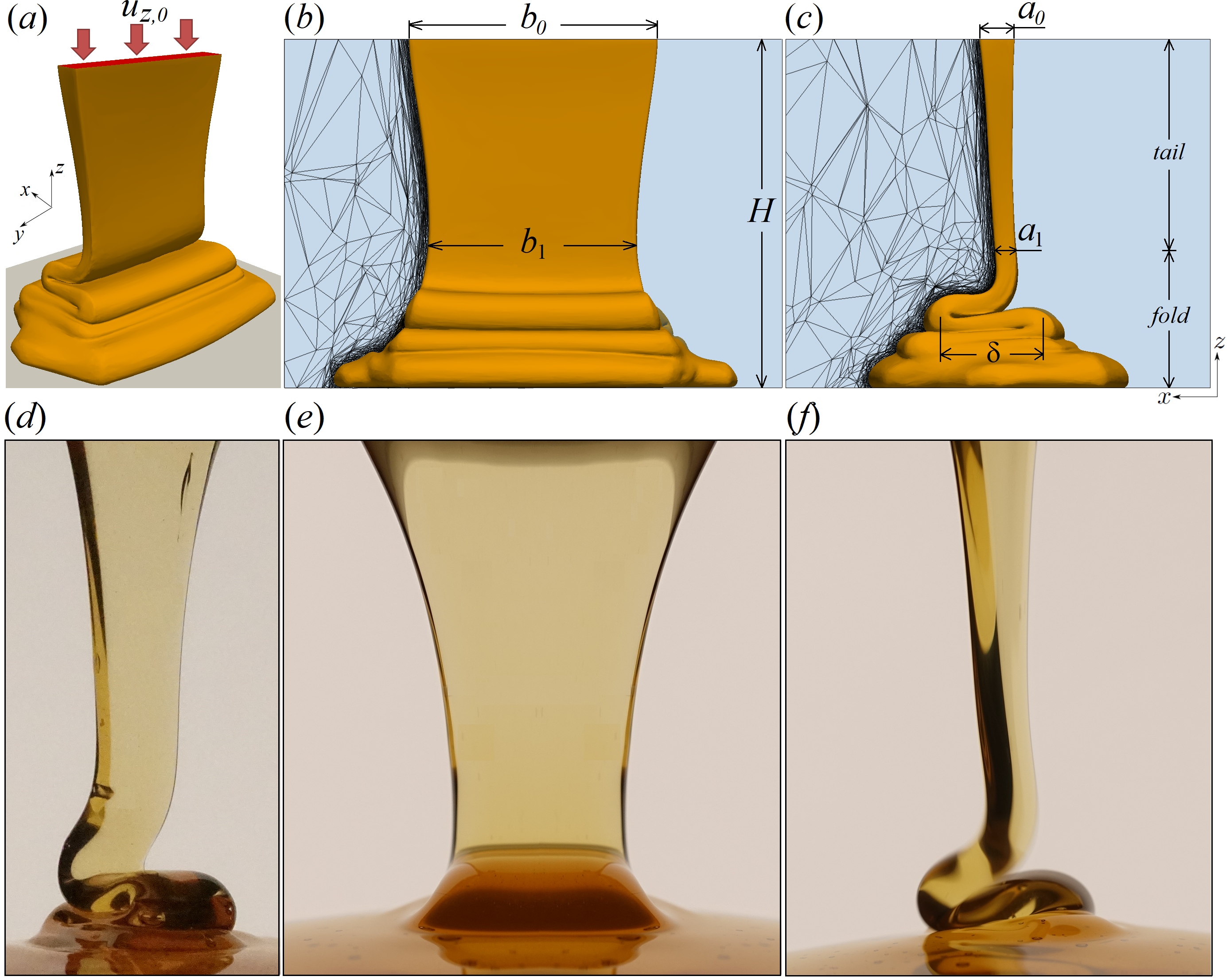}
\vspace*{-0.7cm}
\caption{A power-law fluid (yellow) of density $\rho$ and viscosity $\eta$ leaves extrusion slot of thickness $a_0$ and width $b_0$ with a velocity $u_{z,0}$ and falls onto a solid surface (gray) on which it forms a deformed sheet of amplitude $\delta$ that folds with an folding frequency $\Omega$ in a transversal $x$-$z$ plane. A typical numerical simulation is illustrated in the top line, while a typical experimental one is shown in the bottom one: (\textit{a} and \textit{d}) General view; (\textit{b} and \textit{e}) Frontal view. (\textit{c} and \textit{f}) Side view. The blue part in (\textit{b}) and (\textit{c}) represents the air surrounding the power-law filament. The left-side black lines represent the mesh. A movie related to this figure is available (movie-2).}
\label{fig-2}
\end{figure*}

%--------------------------------------------------------------------------------------------------------------------------------------------------------------------------------------%--------------------------------------------------------------------------------------------------------------------------------------------------------------------------------------
As illustrated in Fig. \ref{fig-2}, we analyse, both numerically (\ref{fig-2}\textit{a}-\textit{c}) and experimentally (Figs. \ref{fig-2}\textit{d}-\textit{f}), the development of folding instabilities in a viscous power-law sheet (yellow part) surrounded by air (blue part). The non-Newtonian fluid of density $\rho$ and viscosity $\eta$ leaves a channel (extrusion slot) of thickness $a_0$ and width $b_0$ with a velocity $u_{z,0}$ and falls onto a solid surface or a fluid substrate on which it forms a folded layer of amplitude $\delta$ at a folding frequency $\Omega$ in a transversal $x$-$z$ plane. The fall height is the distance $H$ from the upper channel to the first point of contact of the free portion of the power-law sheet with the pile of fluid accumulated on the solid surface. The thickness of the trailing part of the sheet generally varies downward and its value in the folded part is $a_1$. The contact point (e.g. upper boundary of the folded part) is located at the $x$-$y$ plane where the sheet vertical velocity $u_z$ reaches its maximum value, in average. Hence at the steady statistically state configuration (periodic folding at constant $\Omega$), the sheet develops a $x$-$y$ averaged vertical velocity $u_{z,1} = Q/(a_1 b_1)$ at the contact $x$-$y$ plane, $Q$ denoting the volume flow rate.

The flow scenario mentioned above is considered through about 400 numerical simulations and 100 experiments. 

%--------------------------------------------------------------------------------------------------------------------------------------------------------------------------------------
\subsection{Physical formulation and Numerical Method}
%--------------------------------------------------------------------------------------------------------------------------------------------------------------------------------------
The computational approach used to simulate the folding process is based on a general solver \citep[CIMLIB-CFD, a parallel, finite element library;][]{Coupez-13} which takes into account the rheological behaviour of each fluid, as well as surface tension effects \citep{Valette-19, Pereira-19b, Valette-20, Pereira-20}. More precisely, the Cauchy stress tensor $\boldsymbol\sigma_c$ is defined as 
\begin{equation}
\boldsymbol{\sigma_c} = -p\boldsymbol{I} + \boldsymbol{\tau} \, ,
\label{eq:sigma-c-tensor}
\end{equation}
where, $p$ is the pressure, $\boldsymbol{I}$ denotes the identity tensor and $\boldsymbol{\tau}$ the \textit{extra stress tensor}. The extra stress tensor is given by 
\begin{equation}
\boldsymbol{\tau} = 2 \eta \boldsymbol{D(u)} \, ,
\label{eq:tau-tensor}
\end{equation}
$\boldsymbol{D(u)}$ representing the rate-of-strain tensor defined as $\boldsymbol{D(u)} = 1/2 \left( \boldsymbol{\nabla u} + \boldsymbol{\nabla u}^T \right)$, and $\boldsymbol{u}$ the velocity vector. The sheet viscosity $\eta$ is computed by using a power-law constitutive model \citep{Ostwald-25}. The latter includes the Papanastasiou regularization \citep[exponential part of the following equation;][]{Papanastasiou-87}: 
\begin{equation}
\eta = k {||\boldsymbol{\dot{\gamma}}||}^{m-1}  \left( 1-e^{-||\boldsymbol{\dot{\gamma}}||/n} \right)^{1-m}   \, ,
\label{eq:eta-e}
\end{equation}
where $k$ is the consistency, $m$ denotes the flow behaviour index, and $\boldsymbol{\dot{\gamma}}$ is twice the rate-of-strain. The norm of $\boldsymbol{\dot{\gamma}}$ is called deformation rate, being defined as $|| \boldsymbol{\dot{\gamma}}|| = \left( \frac{1}{2} \boldsymbol{\dot{\gamma}} : \boldsymbol{\dot{\gamma}} \right) ^{\frac{1}{2}}$ \citep{Bird-87}. Furthermore, $n$ is the Papanastasiou coefficient that allows to bound the value of the viscosity for vanishing $||\boldsymbol{\dot{\gamma}}||$. 

Regarding Eq. \ref{eq:eta-e}, it is important to observe that values of $m$ smaller than 1 ($m < 1$) are related to shear thinning (pseudo-plastic) effects (the viscosity is a decreasing function of $||\boldsymbol{\dot{\gamma}}||$). On the other hand, shear thickening (dilatant) effects emerges if $m > 1$ (the viscosity is an increasing function of $||\boldsymbol{\dot{\gamma}}||$). Finally, a Newtonian behaviour is recovered when $m =1$ (constant viscosity). 

The momentum equation applied to the considered solenoidal flows ($ \boldsymbol{\nabla \cdot u} = 0$) reads: 
\begin{equation}
\rho \left(  \frac{\partial \boldsymbol{u}}{\partial t} + \boldsymbol{u} \cdot \nabla \boldsymbol{u} - \boldsymbol{g} \right) = - \nabla p + \nabla \cdot \boldsymbol{\tau} + \boldsymbol{f_{st}}   \, ,
\label{eq:cons-mom}
\end{equation} 
in which $\rho$, $\nabla$, $\boldsymbol{g}$ , $\nabla \cdot$ and $\boldsymbol{f_{st}}$ are, respectively, the fluid density, the gradient operator, the gravity vector, the divergence operator, and a capillary term related to the surface tension force. Because surface tension has a relatively minor effect on the folding/coiling frequency for typical experimental viscous fluids (between 1\% and 20\%), as pointed out by Ribe et al. \cite{Ribe-12}, we neglect it in the following discussion, i.e. $\boldsymbol{f_{st}}$ is equal to zero.
 
Our numerical methods are based on a Variational Multi-Scale (VMS) approach combined with anisotropic mesh adaptation with highly stretched elements \citep[black lines in Figs. \ref{fig-2}\textit{b} and \ref{fig-2}\textit{c}; ][]{Riber-16, Valette-20}. In order to capture the non-Newtonian fluid/air interface as a function of time, $t$, a level-set method is used \citep{Hachem-16}. Velocity and pressure fields are primitive unknowns that are computed using a unified framework according to which all fluids occupy a single computational mesh by simply mixing the different fluid properties (viscosity, density etc.) and using smoothed Heaviside functions (built from each level-set function) to take property discontinuities into consideration.

The numerical configuration taken into account in this work is illustrated in Figs. \ref{fig-2}(\textit{a}-\textit{c}), where the mesh (composed of approximately $10^6$ elements) is depicted, adapted around each interface (a mesh sensitivity analysis is available in the Supplemental Material). The corresponding zero-isovalues for the level-set function are also shown. A variety of extrusion slots (upper channel) are considered, as indicated across the following Section. In terms of order of magnitude, $ \mathcal{O} \left( 10^{-3} \right) \lesssim a_0 \lesssim \mathcal{O} \left( 10^{-2} \right)$ m and $b_0 \sim \mathcal{O} \left( 10^{-2} \right)$ m. In addition, a wide range of sheet rheological properties ($k$ and $m$), fall height, and gravity is considered: 0.395 $\leq m \leq$ 1.3; $\mathcal{O} \left( 10^0 \right) \lesssim   k  \lesssim   \mathcal{O} \left( 10^2 \right)$ Pa$\cdot$s$^m$; $\mathcal{O} \left( 10^{-3} \right) \lesssim  H \lesssim  \mathcal{O} \left( 10^{0} \right)$ m; $\mathcal{O} \left( 10^{-1} \right) \lesssim  g \lesssim  \mathcal{O} \left( 10^{2} \right)$ m/s$^2$. The viscous fluid densities are kept fixed at $\rho = 1000$ kg/m$^3$. The Papanastasiou coefficient is $n = 10^{-5}$ (see more details in the Supplemental Material). Concerning the air phase, both viscosity $\eta_{air}$ and density $\rho_{air}$ are constant and respectively equal to $10^{-5}$ Pa$\cdot$s and 1 kg/m$^3$. Lastly, initial and boundary conditions for the flow equations are, respectively, initial velocity $u_{z,0}$ at the extrusion slot ($ \leq 0.1$ m/s for all flow cases), and zero normal stress in the air domain.

%--------------------------------------------------------------------------------------------------------------------------------------------------------------------------------------
\subsection{Experimental Procedure}
\begin{figure}%[h!]
\centering
\includegraphics[angle=0, scale=0.21]{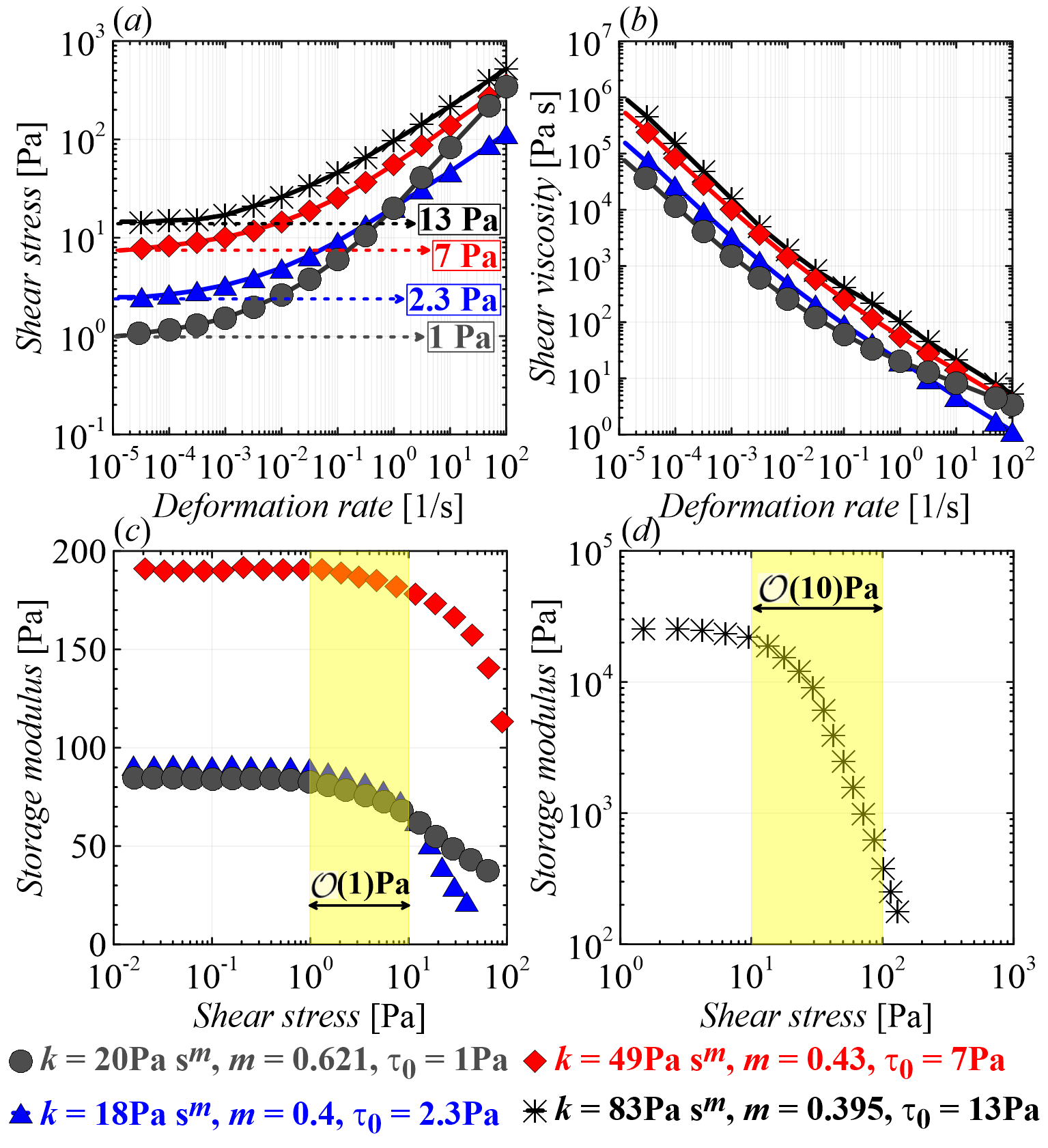}
\vspace*{-0.4cm}
\caption{Rheological behaviour of the considered Carbopol aqueous suspension: (\textit{a}) Shear stress, $||\boldsymbol{\tau}||$, versus deformation rate, $||\boldsymbol{\dot{\gamma}}||$; (\textit{b}) Shear viscosity, $\eta$, versus deformation rate, $||\boldsymbol{\dot{\gamma}}||$; and (\textit{c} and \textit{d}) Storage modulus, $G^{\prime}$, versus shear stress, $||\boldsymbol{\tau}||$. As indicated by the solid lines in (\textit{a}) and (\textit{b}), both shear stress and viscosity data are perfectly fitted by the Herschel-Bulkley equation with the following parameters: $k = 20$ Pa$\cdot$s$^m$, $m = 0.621$, $\tau_0 = 1$ Pa (gray circles); $k = 18$ Pa$\cdot$s$^m$, $m = 0.4$, $\tau_0 = 2.3$ Pa (blue triangles); $k = 49$ Pa$\cdot$s$^m$, $m = 0.43$, $\tau_0 = 7$ Pa (red diamonds); $k = 83$ Pa$\cdot$s$^m$, $m = 0.395$, $\tau_0 = 13$ Pa (black asterisks). The yield stress values obtained by regression analysis and indicated by the boxes in (\textit{a}) are corroborated, in terms of order of magnitude, by the shear stress values related to the initial changes of direction of the $G^{\prime}$ curves, as highlighted by the yellow areas in (\textit{c}) and (\textit{d}).}
\label{fig-3}
\end{figure}

%--------------------------------------------------------------------------------------------------------------------------------------------------------------------------------------
Experiments are also carried out by using five different fluids: a 5.2 Pa.s honey (Newtonian), and four types of Carbopol aqueous suspensions (shear thinning). Some rheological properties of these non-Newtonian materials, measured with an Anton Paar MCR-302 rheometer equipped with a cone-plate geometry, are displayed by Fig. \ref{fig-3}. Each symbol denotes a specific Carbopol type. As indicated by the shear stress, shear viscosity, and storage modulus ($G^{\prime}$) curves displayed in Figs. \ref{fig-3}(\textit{a}-\textit{d}), respectively, the considered suspensions exhibit elasto-viscoplastic ingredients. However, as pointed out in the following lines, both elasticity and plasticity can be neglected for the flow scenarios explored here. 

The solid lines in Fig. \ref{fig-3} indicate that the Carpobol suspension shear stress responses (as well as the viscosity ones) can be perfectly fitted by a simple Herschel-Bulkley equation, $||\boldsymbol{\tau}|| = k {||\boldsymbol{\dot{\gamma}}||}^{m} + \tau_0$,
in which $\tau_0$ denotes a characteristic stress level above which the material moves from a solid-like behaviour to a liquid-like one, i.e. the yield stress. The fit parameters related to each Carpobol suspension are: $k = 20$ Pa$\cdot$s$^m$, $m = 0.621$, $\tau_0 = 1$ Pa (gray circles); $k = 18$ Pa$\cdot$s$^m$, $m = 0.4$, $\tau_0 = 2.3$ Pa (blue triangles); $k = 49$ Pa$\cdot$s$^m$, $m = 0.43$, $\tau_0 = 7$ Pa (red diamonds); $k = 83$ Pa$\cdot$s$^m$, $m = 0.395$, $\tau_0 = 13$ Pa (black asterisks). The yield stress values obtained by regression analysis and stressed by the boxes in Fig. \ref{fig-3}(\textit{a}) are corroborated, in terms of order of magnitude, by the shear stress values related to the initial changes of direction of the $G^{\prime}$ curves \citep{Bird-87}, as highlighted by the yellow areas in Figs. \ref{fig-3}(\textit{c}) and \ref{fig-3}(\textit{d}). 

Using the mentioned Herschel-Bulkley parameters, one can show that the Bingham numbers related to the flow cases considered here (defined as the ratio between the yield stress and the viscous stress, $\textrm{Bn} = \frac{\tau_0}{k(u_{z,1}/a_1)^m}$) are typically lower than 0.1 (with $u_{z,1}/a_1 \geq 10$ s$^{-1}$). Consequently, plastic effects can be neglected. Moreover, comparing the estimated Carbopol relaxation times \citep[$\lambda \approx \left(  k/G^{\prime} \right)^{1/m}$;][]{Luu-09} to the mean folding frequencies globally obtained ($\Omega \leq 10$ s$^{-1}$), one can conclude that elasticity does not play a relevant role either ($\lambda \Omega < 1$). Hence, our Carpobol suspensions can be considered as simple power-law fluids for the flow cases explored in this work, as confirmed by the results discussed in Section \ref{RD}. 

The important properties of the tested materials are summarised in Table \ref{table-1}. Their density is $\rho \approx 1000$ kg/m$^3$. Here, we do not measure their surface tension, $\sigma$, which can be a challenging task when considering yield stress fluids. Instead, we use classically reported measurements as references \citep{Manglik-01, Boujlel-13, Geraud-14, Jorgensen-15}. Honey surface tension usually vary over a range of 50 mN/m and 60 mN/m, and, thus, we assume that $\sigma = 55$ mN/m. For Carbopol aqueous suspensions, the commonly reported values vary between 51 mN/m and 69 mN/m, which leads us to the mean value of $\sigma = 60$ mN/m. As a result, one can show that, for the flow cases explored here, the Capillary number, $\textrm{Ca} = \frac{k \left( u_{z,1}/a_{1} \right)^m + \tau_0}{\sigma/a_{1}}$, is typically larger than 1. Hence, surface tension effects can be neglected.  
%\vspace{0.5cm}
\begin{table}[h]
\caption{Important properties of the tested materials.}
\centering
%\begin{center}
{\footnotesize
\begin{tabular}{lcccccc}
\hline
Sample & $\rho$ [kg/m$^3$] & $k$ [Pa s$^m$] & $m$ & $\tau_0$ [Pa] & $G^{\prime}$ [Pa] & $\sigma$ [N/m] \\
\hline
Honey & 1000 & 5.2 & 1 & 0 & 0 & 0.055\\ %[-0.4cm]
Carbopol 1 & 1000 & 20 & 0.621 & 1 & 85 & 0.06\\ %[-0.4cm]
Carbopol 2 & 1000 & 18 & 0.4 & 2.3 & 91 & 0.06\\ %[-0.4cm]
Carbopol 3 & 1000 & 49 & 0.43 & 7 & 191 & 0.06\\ %[-0.4cm]
Carbopol 4 & 1000 & 83 & 0.395 & 13 & 25403 & 0.06\\ %[-0.4cm]
\hline
\end{tabular} }
%\end{center}
\label{table-1}
\end{table}

Regarding the used experimental apparatus, it is very similar to that described by Habibi et al. \cite{Habibi-14}. Basically, the fluid initially falls at a constant volumetric rate from a large constant-head reservoir, passing through an extrusion slot connected to the bottom of the reservoir, and finally hitting a horizontal plate. The behaviour of the fluid as it strikes the plate is recorded by a high-speed camera [$\mathcal{O}(10^3)$ frames per second] with the aid of a LED backlight panel (see Figs. \ref{fig-2}\textit{d}-\textit{e}). The volumetric rate is determined by the depth of the fluid in the reservoir. Different impact velocities are achieved by varying the distance between the extrusion slot and the hit horizontal plate. Five different extrusion slots are used: $a_0$ x $b_0$ $=$ 0.7 cm x 7 cm; 0.7 cm x 5 cm; 0.5 cm x 4 cm; 0.3 cm x 4 cm; and 0.3 cm x 2.5 cm.   

%%%%%%%%%%%%%%%%%%%%%%%%%%%%%%%%%%%%%%%%%%%%%%%%%%%%%%%%%%%%%%%%%%%%%%%%%%%%%%
%%%%%%%%%%%%%%%%%%%%%%%%%%%%%%%%%%%%%%%%%%%%%%%%%%%%%%%%%%%%%%%%%%%%%%%%%%%%%%
\section{Results and Discussion} \label{RD}
%%%%%%%%%%%%%%%%%%%%%%%%%%%%%%%%%%%%%%%%%%%%%%%%%%%%%%%%%%%%%%%%%%%%%%%%%%%%%%
%%%%%%%%%%%%%%%%%%%%%%%%%%%%%%%%%%%%%%%%%%%%%%%%%%%%%%%%%%%%%%%%%%%%%%%%%%%%%%
%\vspace*{-0.5cm}
As previously reported by Skorobogatiy and Mahadevan \cite{Skorobogatiy-00}, and Ribe \cite{Ribe-03}, in Newtonian viscous sheets, folding instabilities can develop basically two different regimes, each of them corresponding to a different balance among the viscous, gravitational, and inertial forces. In the first regime, the fall height is generally very small and consequently the folding is so slow that both gravity and inertia are negligible when compared with viscous forces. As a result, $a_0 \approx a_1$ and $u_{z,0} \approx u_{z,1}$, which leads to a folding amplitude and a folding frequency simply given by $\delta \sim H$ and $\Omega \sim u_{z,1}/H$, respectively. The second regime is often observed at higher fall heights, when the fluid falls freely after left the extrusion slot. It emerges from a balance between gravity and viscous forces of which $\delta$ is a decreasing function, while $\Omega$ exhibits an opposite behaviour. In other words, when the sheet is highly exposed to gravity forces, one observes the formation of smaller folds at higher frequencies. Eventually, $\delta$ becomes so small that folding ceases and, thus, the fluid simply spreads after hitting the solid surface.  

In the present Section, we extend the Newtonian analyses mentioned above by highlighting shear thinning/thickening effects not only on the referred \textit{viscous} and \textit{gravitational} folding regimes (Subsection \ref{FR}), but also on the \textit{folding suppression} (Subsection \ref{FC}).   

%%%%%%%%%%%%%%%%%%%%%%%%%%%%%%%%%%%%%%%%%%%%%%%%%%%%%%%%%%%%%%%%%%%%%%%%%%%%%%
\subsection{Folding regimes} \label{FR}
%%%%%%%%%%%%%%%%%%%%%%%%%%%%%%%%%%%%%%%%%%%%%%%%%%%%%%%%%%%%%%%%%%%%%%%%%%%%%%

\begin{figure*}%[h!]
\centering
\includegraphics[angle=0, scale=0.275]{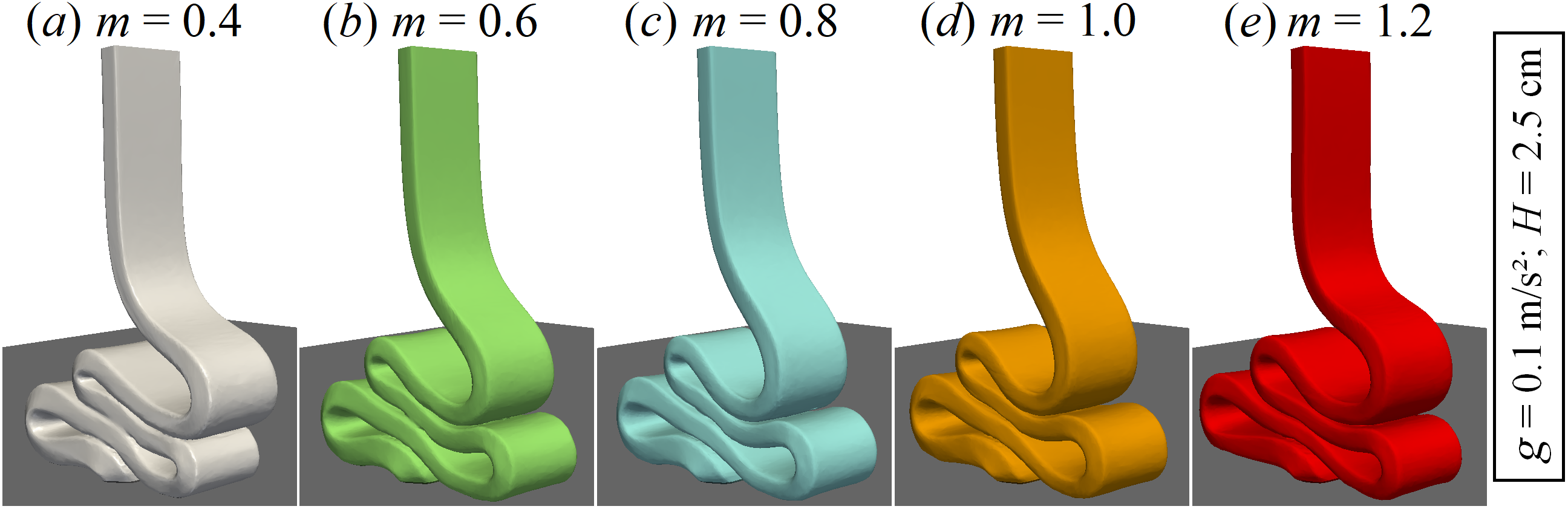}
\vspace*{-0.7cm}
\caption{Instantaneous snapshots (taken at the same instant $t$) of folding instabilities that emerge from different power-law sheets. The dimensions of the extrusion slot are 0.14 cm x 0.5 cm. The fluid initial vertical velocity ($u_{z,0} = 0.1$ m/s), the consistency ($k = 10$ Pa$\cdot$s$^m$), the fluid density ($\rho = 1000$ kg/m$^3$), the fall height ($H = 2.5$ cm) and gravity ($g = 0.1$ m/s$^2$) are kept fixed, while five flow behaviour indexes are considered: $m = 0.4,~0.6,~0.8$ (shear-thinning fluids); $m = 1$ (Newtonian fluid); $m = 1.2$ (shear-thickening fluids). These non-Newtonian filaments develop similar folding instabilities. This process is also shown in a supplemental movie (movie-3).}
\label{fig-4}
%\vspace*{-0.5cm}
\end{figure*}  

\begin{figure*}%[h!]
\centering
\includegraphics[angle=0, scale=0.19]{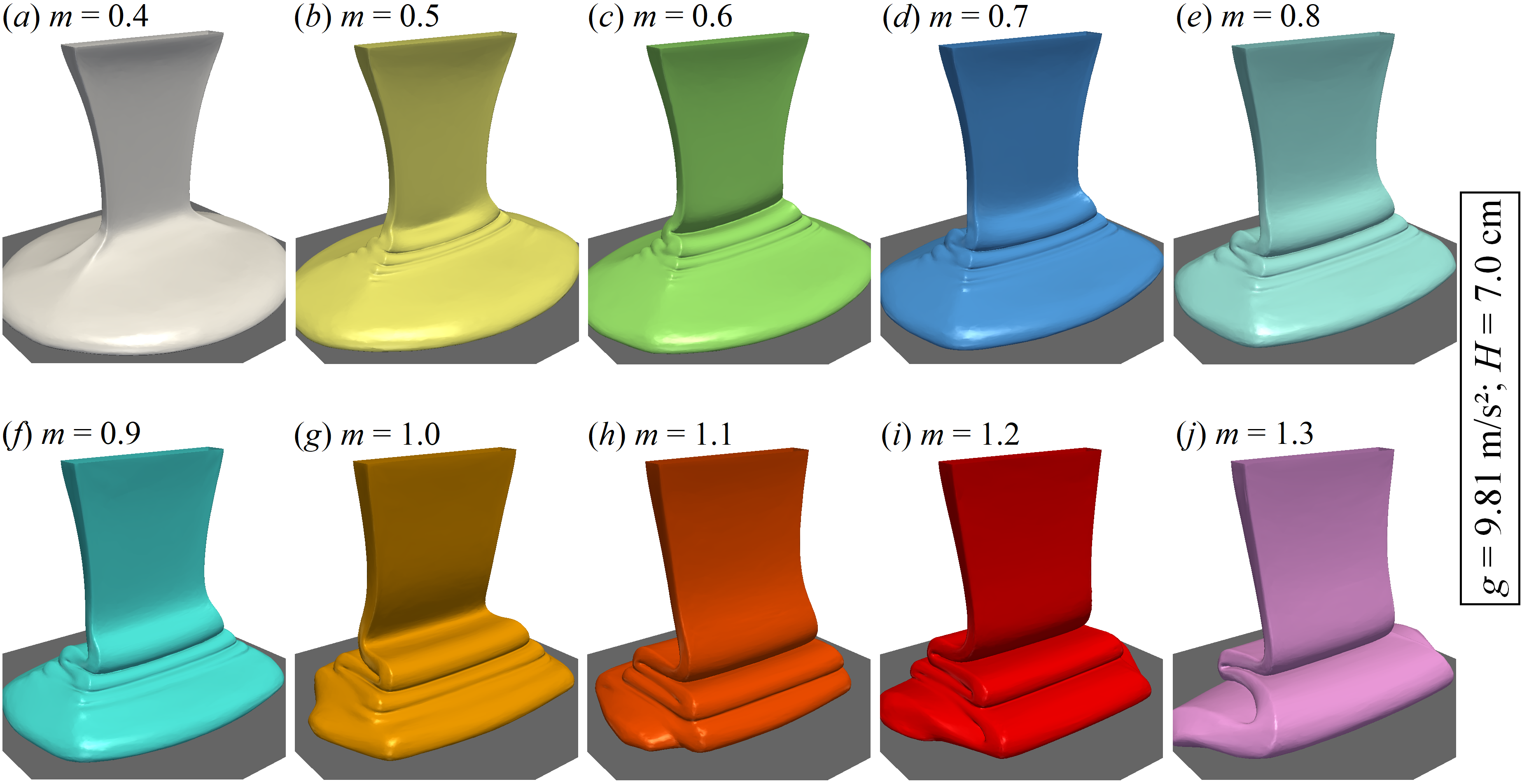}
\vspace*{-0.4cm}
\caption{Instantaneous snapshots (taken at the same instant $t$) of folding instabilities that emerge from different power-law fluids. The dimensions of the extrusion slot are 0.7 cm x 5.0 cm. The fluid initial vertical velocity ($u_{z,0} = 0.1$ m/s), the consistency ($k = 10$ Pa$\cdot$s$^m$), the fluid density ($\rho = 1000$ kg/m$^3$), the fall height ($H = 7.0$ cm) and gravity ($g = 9.81$ m/s$^2$) are kept fixed, while ten flow behaviour indexes are considered: $m = 0.4,~0.5,~0.6,~0.7,~0.8,~0.9$ (shear-thinning fluids); $m = 1$ (Newtonian fluid); $m = 1.1,~1.2,~1.3$ (shear-thickening fluids). This process is also shown in a supplemental movie (movie-4).}
\label{fig-5}
\vspace*{-0.5cm}
\end{figure*}  

Initially, we focus on small fall height at low gravity, such those illustrated by Fig. \ref{fig-4} for which $H = 2.5$ cm and $g = 0.1$ m/s$^2$. This figure shows instantaneous snapshots (taken at the same instant $t$) of folding instabilities in different power-law sheets. The initial fluid vertical velocity ($u_{z,0} = 0.1$ m/s), the consistency ($k = 10$ Pa$\cdot$s$^m$), the fluid density ($\rho = 1000$ kg/m$^3$), and the extrusion slot dimensions ($a_0 = 0.14$ cm and $b_0 = 0.5$ cm) are kept fixed, while five flow behaviour indexes are considered: $m = 0.4,~0.6,~0.8$ (shear-thinning fluids); $m = 1$ (Newtonian fluid); $m = 1.2$ (shear-thickening fluids). Despite their rheological differences, these sheets develop similar folding instabilities in terms of both frequency and amplitude, which indicates that their non-Newtonian nature plays no role in the folding dynamics. 

A different scenario from that pictured above emerges by accentuating gravitational effects ($\propto \rho g H$) through the increase of both $H$ and $g$, as shown by Fig. \ref{fig-5}. In this figure, for which $H = 7.0$ cm and $g = 9.81$ m/s$^2$, instantaneous snapshots (taken at the same instant $t$) of folding instabilities in different power-law sheets are considered. The fluid initial vertical velocity ($u_{z,0} = 0.1$ m/s), the consistency ($k = 10$ Pa$\cdot$s$^m$), the fluid density ($\rho = 1000$ kg/m$^3$), and the extrusion slot dimensions ($a_0 = 0.7$ cm and $b_0 = 5.0$ cm) are kept fixed, while ten flow behaviour indexes are considered: $m = 0.4,~0.5,~0.6,~0.7,~0.8,~0.9$ (shear-thinning fluids); $m=1$ (Newtonian fluid); $m = 1.1,~1.2,~1.3$ (shear-thickening fluids). Clearly, shear-thickening fluids ($m > 1$) exhibit more pronounced folding amplitudes, as well as lower folding frequencies (only three folded layers were formed at $m = 1.3$ against five of them at $m = 1.2$, for example). Moreover, the instabilities can eventually be suppressed by high shear thinning effects, as displayed in Fig. \ref{fig-5}(\textit{a}) for which $m = 0.4$ and the fluid simply spreads after the impact (the folding suppression will be discussed in details in Subsection \ref{FC}). 

\begin{figure}%[h!]
\centering
\includegraphics[angle=0, scale=0.24]{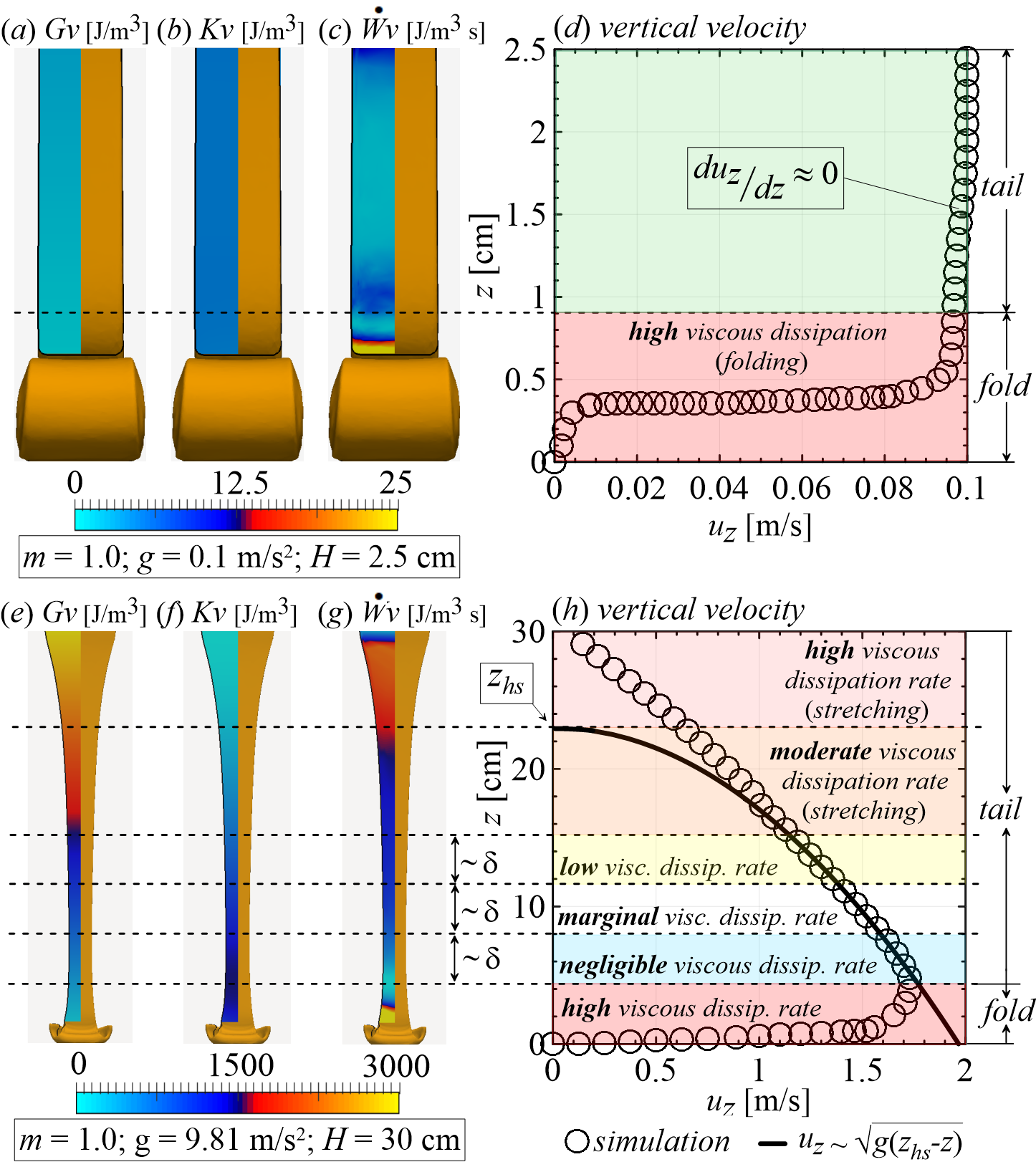}
\vspace*{-0.7cm}
\caption{Folding instabilities obtained for the same material ($k=$ 10 Pa$\cdot$s$^m$, $m=$ 1.0, $\rho=$ 1000 kg/m$^3$) at two $H$-$g$ couples: $H = 2.5$ cm and $g = 0.1 $ m/s$^2$(\textit{a}-\textit{d}; $a_0 = 0.14$ cm and $b_0 = 0.5$ cm); and $H = 30$ cm and $g = 9.81$ m/s$^2$ (\textit{e}-\textit{h}; $a_0 = 0.7$ cm and $b_0 = 5.0$ cm). For each of them, we consider the contours of the gravitational potential energy per unit volume $G_v$, the kinetic energy per unit volume $K_v$, and the viscous dissipation rate per unit volume ${\dot{W}}_v$ along the centre $y$-$z$ plane (left-side plane in \textit{a}-\textit{c} and \textit{e}-\textit{g}; the yellow parts on both the right-side and the bottom of \textit{a}-\textit{c} and \textit{e}-\textit{g} illustrate the sheet surfaces), as well as the $x$-$y$ average of the vertical velocity along the sheet height $z$ (gray circles in \textit{d} and \textit{h}). The color code for the coloured areas in (\textit{d}) and (\textit{h}) is which follows: green $\Rightarrow$ $du{z}/dz \approx 0$; red $\Rightarrow$ high viscous dissipation by folding; pink $\Rightarrow$ high viscous dissipation by stretching; orange $\Rightarrow$ moderate viscous dissipation by stretching; yellow $\Rightarrow$ low viscous dissipation; white $\Rightarrow$ marginal viscous dissipation; blue $\Rightarrow$ negligible viscous dissipation.}
\label{fig-6}
\vspace*{-0.5cm}
\end{figure}

Comparing Figs. \ref{fig-4} and \ref{fig-5}, one can formulate that the physical contrasts between these two different folding dynamics rise from gravitational effects ($\propto \rho g H$). In order to better understand these distinguishing behaviours, we take into account in Fig. \ref{fig-6} folding instabilities observed for the same material ($k = 10$ Pa$\cdot$s$^m$, $m = 1.0$, $\rho = 1000$ kg/m$^3$) at two $H$-$g$ couples: $H = 2.5$ cm and $g = 0.1$ m/s$^2$ (upper-line results, Figs. \ref{fig-6}\textit{a}-\textit{d}; $\rho g H = 2.5$ Pa); $H = 30$ cm and $g = 9.81$ m/s$^2$ (bottom-line results, Figs. \ref{fig-6}\textit{e}-\textit{h}; $\rho g H = 2943$ Pa). For each of them, we display the contours of the gravitational potential energy per unit volume $G_v$ ($= \rho g z$), the kinetic energy per unit volume $K_v$ ($= \rho {||\boldsymbol{u}||}^2/2$), and the viscous dissipation rate per unit volume ${\dot{W}}_v$ ($=k {||\boldsymbol{\dot{\gamma}}||}^{m+1}$) along the centre $y-z$ plane (left-side planes in Figs. \ref{fig-6}\textit{a}-\textit{c} and Figs. \ref{fig-6}\textit{e}-\textit{g}; the yellow three dimensional structures on both the right-side and the bottom of Figs. \ref{fig-6}\textit{a}-\textit{c} and Figs. \ref{fig-6}\textit{e}-\textit{g} illustrate the sheet surfaces), as well as the $x-y$ average of the vertical velocity $u_z$ along the sheet height $z$ (gray circles in Figs. \ref{fig-6}\textit{d} and \ref{fig-6}\textit{h}). 

At $H = 2.5$ cm and $g = 0.1$ m/s$^2$ (Figs. \ref{fig-6}\textit{a}-\textit{d}), negligible variations of $G_v$, $K_v$ and ${\dot{W}}_v$ are observed along the tail. As a result, no change in the sheet dimensions occurs during the fall ($a_0 \approx a_1$, and $b_0 \approx b_1$) and the vertical velocity $u_z$ remains constant throughout the tail ($u_{z,0} \approx u_{z,1}$; $du_{z}/dz \approx 0$), as indicated by the gray circles in the green region of Fig.\ref{fig-6}(\textit{d}). However, the sheet velocity quickly decreases towards zero along the fold (as pointed out by the red region in Fig. \ref{fig-6}\textit{d}), where the viscous dissipation rate becomes pronounced (see the contours of ${\dot{W}}_v$). Lastly, it is important to emphasise that, for the refereed case, both gravitational and inertial forces are negligible when compared to the viscous ones, i.e $\frac{\rho g H}{k (u_{z,1}/a_1)^m} \sim \mathcal{O} \left( 10^{-3} \right)$ and $\frac{\rho u_{z,1}^2}{k (u_{z,1}/a_1)^m} \sim  \mathcal{O} \left( 10^{-2} \right)$.

A more complex scenario emerges at $H = 30$ cm and $g = 9.81$ m/s$^2$ (Figs.\ref{fig-6}\textit{e}-\textit{h}). In this case, the sheet tail is highly exposed to gravitational stresses ($\propto gH$). Hence, below the extrusion slot, the sheet velocity tends to increase. More specifically, in the vicinities of the upper channel (pink area), where significant changes in the sheet dimensions occur, the gravity-induced stretching is counterbalanced by viscous forces. As a result, a high viscous dissipation rate is observed in this region. However, below a critical height $z_{hs} \approx 23$ cm (stressed by the black box in Fig. \ref{fig-6}\textit{h}), viscous dissipation rate decreases, while gravity starts to play a more significant role (orange and yellow areas). The vertical velocity $u_z$ tends then to $u_{z} \sim \sqrt{g~(z_{hs}-z)}$ as indicated parabolic fit illustrated by the black solid line. A little below, within both the white and the blue areas of height $\sim \delta$, viscous dissipation almost vanishes and, consequently, the material falls freely, reaching its maximum velocity at the upper boundary of the fold part of the sheet. Since the gravitational potential energy is almost fully converted into kinetic energy along bottom-half of the tail, its vertical velocity variation scales as $\Delta u_{z} \sim \sqrt{g ~ \delta}$. In other words, within both the white and the blue areas, gravitational potential energy is primarily converted into kinetic energy, as noticed by comparing the correspondent contours in Figs. \ref{fig-6}\textit{e} and \ref{fig-6}\textit{f} (the variation of $G_v$ is comparable to that of $K_v$, in terms of order of magnitude, along the referred areas). Nevertheless, after achieving its maximum value (at the interface between the blue and the red area), the fluid kinetic energy is dissipated by viscous effects in the folded part (red area in Fig. \ref{fig-6}\textit{h}), where the sheet of $a_1 \sim \mathcal{O} \left( 10^{-3} \right)$ m is under compression stresses. Within this part, ${\dot{W}}_v$ reaches its higher values and $u_{z}$ is a decreasing function of $H-z$ (gray circles). %Finally, concerning this particular folding case, it is worth noticing that both gravitational and inertial effects are equally relevant, since $\frac{\rho g H}{k (u_{z,1}/a_1)^m} \sim (\mathcal{O})10^{-1}$ and $\frac{\rho u_{z,1}^2}{k (u_{z,1}/a_1)^m} \sim (\mathcal{O})10^{-1}$.      

\begin{figure*}%[h!]
\centering
\includegraphics[angle=0, scale=0.33]{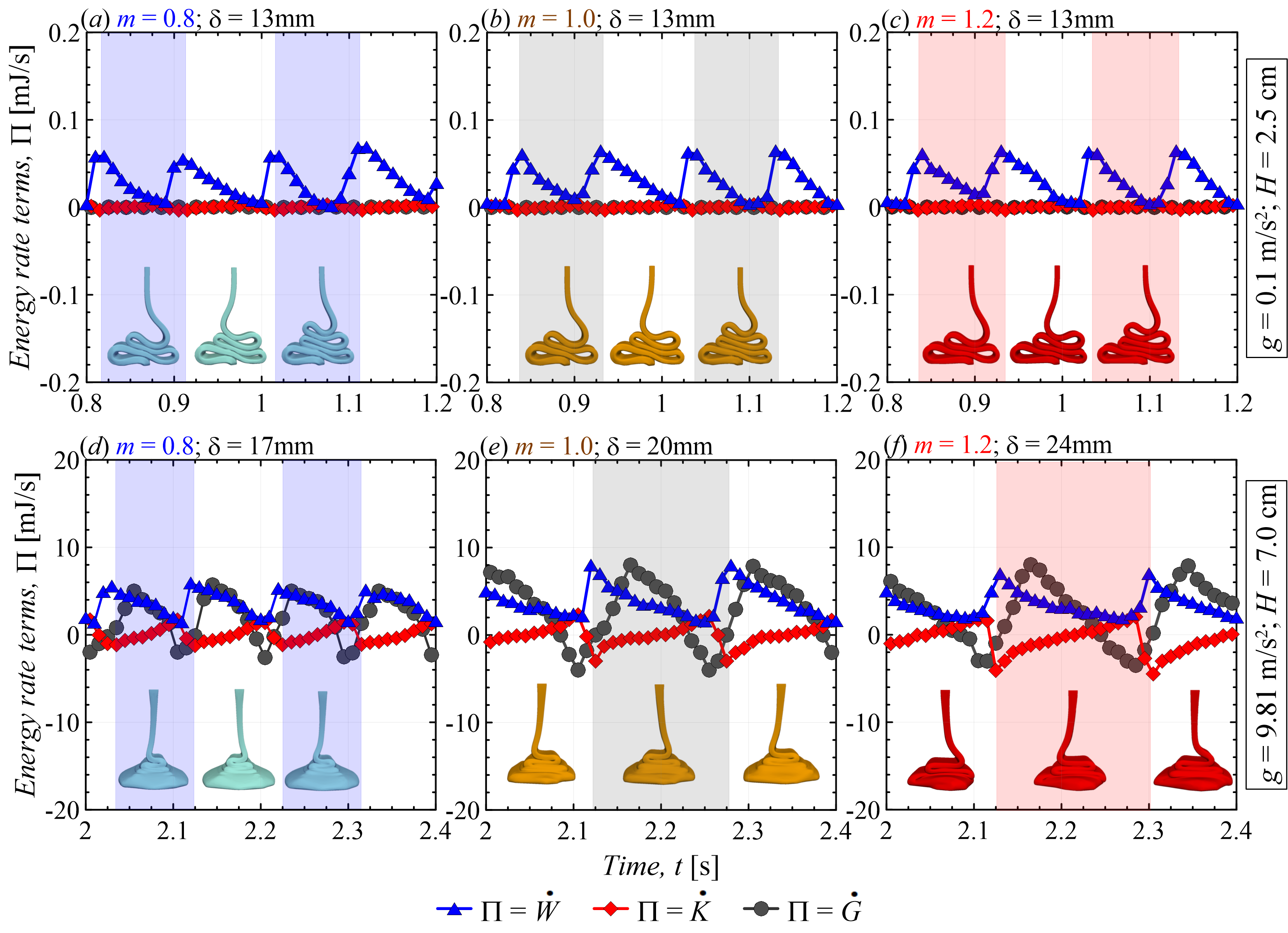}
\vspace*{-0.7cm}
\caption{Viscous dissipation $\dot{W}$ (blue triangles), kinetic energy $\dot{K}$ (red diamonds), and gravitational potential energy $\dot{G}$ (gray circles) rate terms are displayed as a function of time, $t$, for three power-law fluids at $H= 2.5$ cm (upper line; $a_0 = 0.14$ cm and $b_0 = 0.5$ cm) and $H= 7.0$ cm (bottom line; $a_0 = 0.7$ cm and $b_0 = 5.0$ cm): (\textit{a} and \textit{d}) $m=0.8$; (\textit{b} and \textit{e}) $m=1.0$; (\textit{c} and \textit{f}) $m=1.2$. The initial fluid vertical velocity $u_{z,0}$ (0.1 m/s), the consistency $k$ (10 Pa$\cdot$s$^m$), the fluid density (1000 kg/m$^3$), and gravity $g$ (9.81 m/s$^2$) are kept fixed. The width of blue, gray and red regions indicates the time required to form a half-fold. The energy terms are calculated by considering only the fold part of each sheet (i.e. tails are not considered).}
\label{fig-7}
\vspace*{-0.35cm}
\end{figure*} 

Figure \ref{fig-6} allows us to anticipate the existence of at least two folding dynamics. In the first one, observed when both gravitational and inertial effects are negligible, the flow is dominated by viscous forces ($\nabla \cdot \boldsymbol{\tau}=0$, according to the moment equation) that minimize deformations along the tail. Consequently, $a_0 \approx a_1$ and $u_{z,0} \approx u_{z,1}$. Following its impact on the solid bottom surface, the formed sheet of height $H$ is compressed at a constant velocity $u_{z,0} \approx u_{z,1}$ due the uninterrupted injection of material. Since the energy related to the folding deformation [$\propto (u_{z,1}/H)(a_0/H)$] is smaller than the cost of compression ($\propto u_{z,1}/H$), the viscous sheet buckles, generating a fold of amplitude $\delta \sim H$ at a frequency $\Omega \sim u_{z,1}/H$ \citep[see also][]{Ribe-03, Pereira-19b}. Hence, in this scenario called here \textit{viscous regime}, the folding dynamics is completely disconnected from non-Newtonian aspects of the material, as previously indicated in Fig. \ref{fig-4}, and 
\begin{equation}
\Omega^{*}=\frac{\Omega}{u_{z,1}/H}  \sim 1  \, ,
\label{eq: omega_fold_v}
\end{equation} 
\begin{equation} 
\delta^{*}=\frac{\delta}{H} \sim  1   \, . 
\label{eq: delta_fold_v} 
\end{equation}
In other words, in the \textit{viscous regime}, both the dimensionless frequency $\Omega/(u_{z,1}/H)$ and amplitude $\delta/H$ are constant and viscosity independent. These results are rather in line with those reported by Ribe \cite{Ribe-03} by considering Newtonian viscous fluids. The second scenario is related to gravity-exposed sheets and called here \textit{gravitational regime}. In the upper part of the sheet (pink and orange areas in Fig. \ref{fig-6}\textit{h}), gravitational potential energy $G$ is primarily dissipated by viscous effects during the stretching process, while in its bottom part (white, blue, red areas Fig. \ref{fig-6}\textit{h}), the folding dynamics is driven by an energy cascade through which $G$ is converted into kinetic energy $K$ (in both white and blue areas) before being finally dissipated by viscous effects $W$ in the folded part (red area). As a result, time-variations of the referred energy terms within the bottom part of the sheet become comparable with each other ($\dot{G} \sim \dot{K} \sim \dot{W}$). Such energy dissipation mechanism suggests that the gravitational regime is, indeed, affected by the fluid rheology. %Hence, since the folding dynamics in the \textit{gravitational regime} would be driven by a energy cascade linking $G$, $K$, and $W$, one can naturally imagine that it is affected by the fluid rheology.    

The energy transfer arguments anticipated above are confirmed by Fig. \ref{fig-7}, in which viscous dissipation $\dot{W}$, kinetic energy $\dot{K}$, and gravitational potential energy $\dot{G}$ rate terms are displayed as a function of time $t$ for three power-law sheets of $\rho = 1000$ kg/m$^3$ and $k = 10$ Pa$\cdot$s$^m$: $m = 0.8$ (Figs. \ref{fig-7}\textit{a} and \ref{fig-7}\textit{d}); $m = 1.0$ (Figs. \ref{fig-7}\textit{b} and \ref{fig-7}\textit{e}); and $m = 1.2$ (Figs. \ref{fig-7}\textit{c} and \ref{fig-7}\textit{f}). These sheets are submitted to two different $H$-$g$ couples: $H = 2.5$ cm and $g = 0.1$ m/s$^2$ (upper line); and $H = 7.0$ cm and $g = 9.81$ m/s$^2$ (bottom line). The referred energy terms are calculated taking into account only the fold part of the sheets (i.e. their tails are not considered), and they are defined as: 
\begin{equation}
\dot{W} = \int_{V}^{} \eta {||\boldsymbol{\dot{\gamma}}||}^2 dV ~~ \textrm{(viscous dissipation rate, J/s)} \, ,
\label{eq:W} 
\end{equation}
\begin{equation}
\dot{K} = \frac{1}{2}\frac{\partial{\int_{V}^{} \rho {\parallel \boldsymbol{u} \parallel}^2 dV}}{\partial{t}} ~~ \textrm{(kinetic energy rate, J/s)} \, ,
\label{eq:K} 
\end{equation}  
and
\begin{equation}
\dot{G} = \frac{\partial{\int_{V}^{} \rho g z dV}}{\partial{t}} ~~ \textrm{(gravitational potential energy rate, J/s)} \, ,
\label{eq:G} 
\end{equation}  
$V$ being the volume of the fold part of the sheet. Blue, gray, red and white regions indicate the formation of a half-fold (in other words, a single folded layer). The width of these regions represents, thus, the time required to form a fold. Lastly, an interval of  $\Delta t = 0.4$ s is considered. 

Clearly, at $H = 2.5$ cm and $g = 0.1$ m/s$^2$ (upper-line cases), both $\dot{G}$ (gray circles) and $\dot{K}$ (red diamonds) are negligible. Thus, the folding dynamics is basically dominated by the viscous forces related to the dissipation of the compression energy introduced into the system due to the uninterrupted injection of material. Since the non-Newtonian aspects of the considered power-law fluids play no role in the \textit{viscous regime}, similar energy rate curves are exhibited by the upper-line cases. Furthermore, folded layers with similar amplitude and frequency are formed, as observed by comparing the width of the upper coloured areas ($\delta = 1.3$ cm and, thus, comparable to $H$ in terms of order of magnitude; $\Omega \approx 1/0.2$ s$^{-1}$). A contrasting folding dynamics emerges at $H = 7.0$ cm and $g = 9.81$ m/s$^2$ (bottom line), since $\dot{G} \approx \dot{K} \approx \dot{W}$, a direct consequence of the energy cascade discussed in the previous paragraph. The bottom-line results allow us to stress scaling laws based on a global energetic approach, as shown in the following lines.

\begin{figure}%[h!]
\centering
\vspace*{-0.2cm}
\includegraphics[angle=0, scale=0.52]{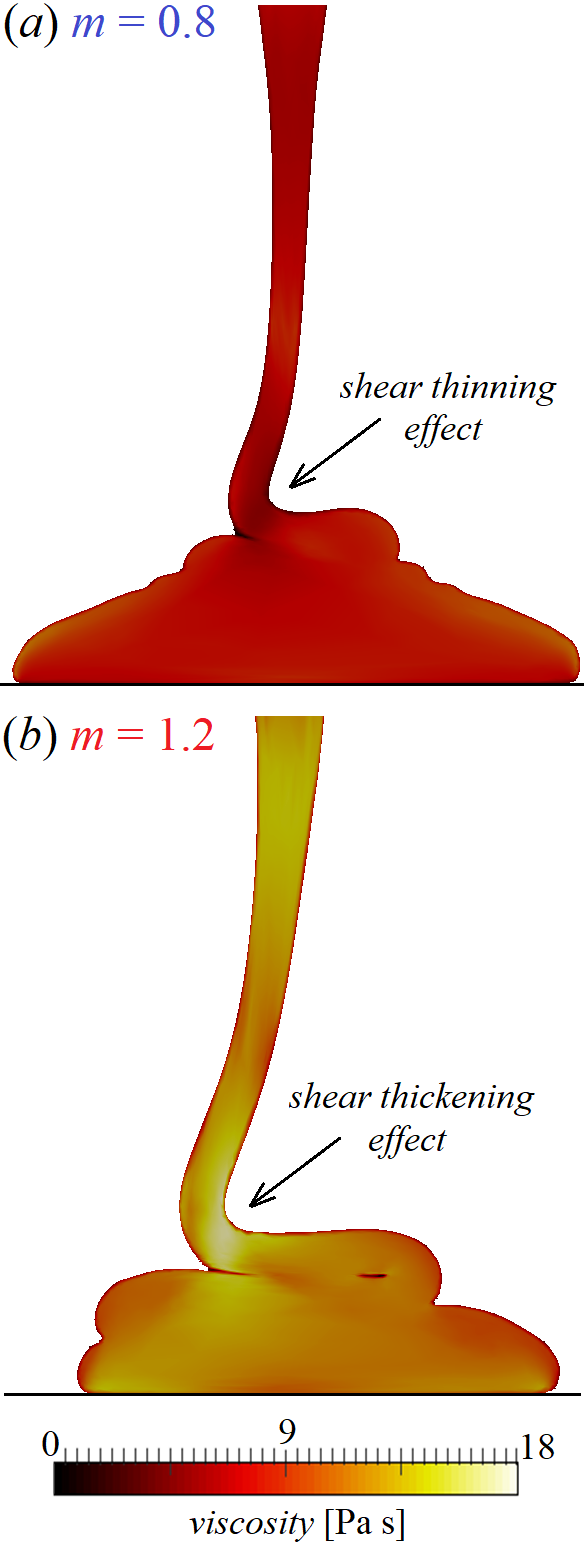}
\vspace*{-0.5cm}
\caption{Viscosity contours displayed in the folding $x$-$z$ plane for two power-law cases: (\textit{a}) $m = 0.8$; (\textit{b}) $m = 1.2$. An extrusion slot with $a_0 = 0.7$ cm and $b_0 = 5.0$ cm is used. The initial fluid vertical velocity ($u_{z,0} = 0.1$ m/s), the consistency ($k = 10$ Pa$\cdot$s$^m$), the fluid density ($\rho = 1000$ kg/m$^3$) the fall height ($H = 7.0$ cm), and gravity ($g = 9.81$ m/s$^2$) are kept fixed. Both shear thinning and shear thickening effects are indicated by the black arrows.}
\label{fig-8}
\end{figure}

We focus on a filament portion of volume $V_{1,f} \sim a_1 b_1 \delta$ situated in the bottom-half of the tail (for example, within the yellow region highlighted in Fig. \ref{fig-6}). It falls of a height $\sim \delta$ to form a single folded layer of length $\sim \delta$ in the fold part of the sheet. As mentioned previously, during its fall, its gravitational potential energy ${G}_{1,f}$ is initially converted into kinetic energy ${K}_{1,f}$ before being dissipated by viscous effect in fold part of the sheet ${W}_{1,f}$. Consequently, $\dot{G}_{1,f} \sim \dot{K}_{1,f} \sim \dot{W}_{1,f}$, or, more directly, $\dot{G}_{1,f} \sim \dot{W}_{1,f}$.  

The gravitational potential energy variation $\dot{G}_{1,f}$ during the formation of a single fold of dimensions $a_1$, $b_1$, and $\delta$ can be expressed as
\begin{equation}
\dot{G}_{1,f} = \frac{\Delta G_{1,f}}{\Delta t} \sim m_{1,f} ~ g ~ \delta ~ \Omega \, ,
\label{eq: G_point}
\end{equation}
where its mass $m_{1,f}$ is $m_{1,f} \sim \rho \delta a_1 b_1$. Hence, 
\begin{equation}
\dot{G}_{1,f} \sim  \rho ~ g ~ \delta^2 a_1 b_1 ~ \Omega \, .
\label{eq: G_point_2}
\end{equation}
In addition, the variations of viscous dissipation energy variation $\dot{W}_{1,f}$ during the folding process is defined as
\begin{equation}
\dot{W}_{1,f} = \int_{V_{1,f}} \eta \dot{\gamma}_{1,f}^2 ~ {dV}_{1,f} \sim \int_{a_1}^{} k {\dot{\gamma}}_{1,f}^{m+1} \delta b_1 ~ da_1 \, ,
\label{eq: W_point}
\end{equation}
in which $V_{1,f}$ denotes the volume of a fold, and ${\dot{\gamma}}_{1,f}$ represents the characteristic bending deformation rate. The latter can be simply expressed as 
\begin{equation}
{\dot{\gamma}}_{1,f} = \frac{\Delta \gamma_{1,f}}{\Delta t} \sim \frac{a_1}{\delta} ~ \Omega \, .
\label{eq: gamma_point}
\end{equation}
Consequently, an expression for $\dot{W}_{1,f}$ is found by replacing Eq. \ref{eq: gamma_point} in Eq.\ref{eq: W_point}:
\begin{equation}
\dot{W}_{1,f} \sim \frac{k ~ b_1 ~ a_1^{m+2} ~ \Omega^{m+1}}{\delta^m} \, .
\label{eq: W_point_2}
\end{equation}
Assuming that $\dot{G}_{1,f} \sim \dot{W}_{1,f}$ (as discussed above), and that $u_{z,1} \sim \delta \Omega$ by mass conservation, we find the folding frequency 
\begin{equation}
\Omega \sim (\frac{u_{z,1}^{m+2}\rho g}{k a_1^{m+1}})^{\frac{1}{2m+2}} \sim \frac{u_{z,1}}{a_1} \left[ \frac{\rho g a_1}{k \left( u_{z,1} / a_{1} \right)^m}  \right]^{\frac{1}{2m+2}} \, ,
\label{eq: omega_fold_g1}
\end{equation}
and the folding amplitude 
\begin{equation}
\delta \sim \left(  \frac{k ~ u_{z,1}^{m} ~ a_{1}^{m+1}}{\rho g} \right)^{\frac{1}{2m+2}} \sim a_{1} \left[  \frac{k \left( u_{z,1} / a_{1} \right)^m}{\rho g a_{1}} \right]^{\frac{1}{2m+2}} \, .
\label{eq: delta_fold_g1}
\end{equation}
Finally, rewriting both Eqs. \ref{eq: omega_fold_g1} and \ref{eq: delta_fold_g1} as a function of the impact Galileo number defined as the ratio between gravity and viscous forces, $\textrm{Ga} = \frac{\rho g a_1}{k(u_{z,1}/a_1)^m}$: 
\begin{equation}
\Omega^{*}=\frac{\Omega}{u_{z,1}/H}  \sim  \frac{H}{a_1} \textrm{Ga}^{\frac{1}{2m+2}}  \, ,
\label{eq: omega_fold_g2}
\end{equation} 
\begin{equation} 
\delta^{*}=\frac{\delta}{H} \sim \frac{1}{(H/a_1) ~ \textrm{Ga}^{\frac{1}{2m+2}}}   \, . 
\label{eq: delta_fold_g2} 
\end{equation}

Equations \ref{eq: omega_fold_g2} and \ref{eq: delta_fold_g2} stress a folding process that emerges from the balance between gravity and viscous forces. We name it \textit{gravitational regime}. Moreover, Eqs. \ref{eq: omega_fold_g2} and \ref{eq: delta_fold_g2} indicate that, at a fixed sheet slenderness and Galileo number, the dimensionless folding frequency decreases with increasing $m$, while the dimensionless folding amplitude becomes more pronounced. This tendency is corroborated not only by Fig. \ref{fig-5}, but also by the increase of the bottom coloured areas with increasing $m$ in Fig. \ref{fig-7}. Such a behaviour is related to the fact that shear thickening materials develop higher viscosity values within the folded parts, as indicated by the viscosity contours displayed in the folding $x$-$z$ plane in Fig. \ref{fig-8}. More specifically, because of the development of larger viscous forces within the folded parts of the shear-thickening filament shown in Fig. \ref{fig-8}(\textit{b}) (lighted part indicated by the black arrow), the latter exhibits a more pronounced resistance to deformation and hence a larger folding amplitude than the shear-thinning fluid illustrated by Fig. \ref{fig-8}(\textit{a}). The latter, in contrast, exhibits smaller viscosity values in the buckled part (see the dark regions pointed out by the black arrow in Fig. \ref{fig-8}\textit{a}).       

\begin{figure}%[h!]
\centering
\includegraphics[angle=0, scale=0.45]{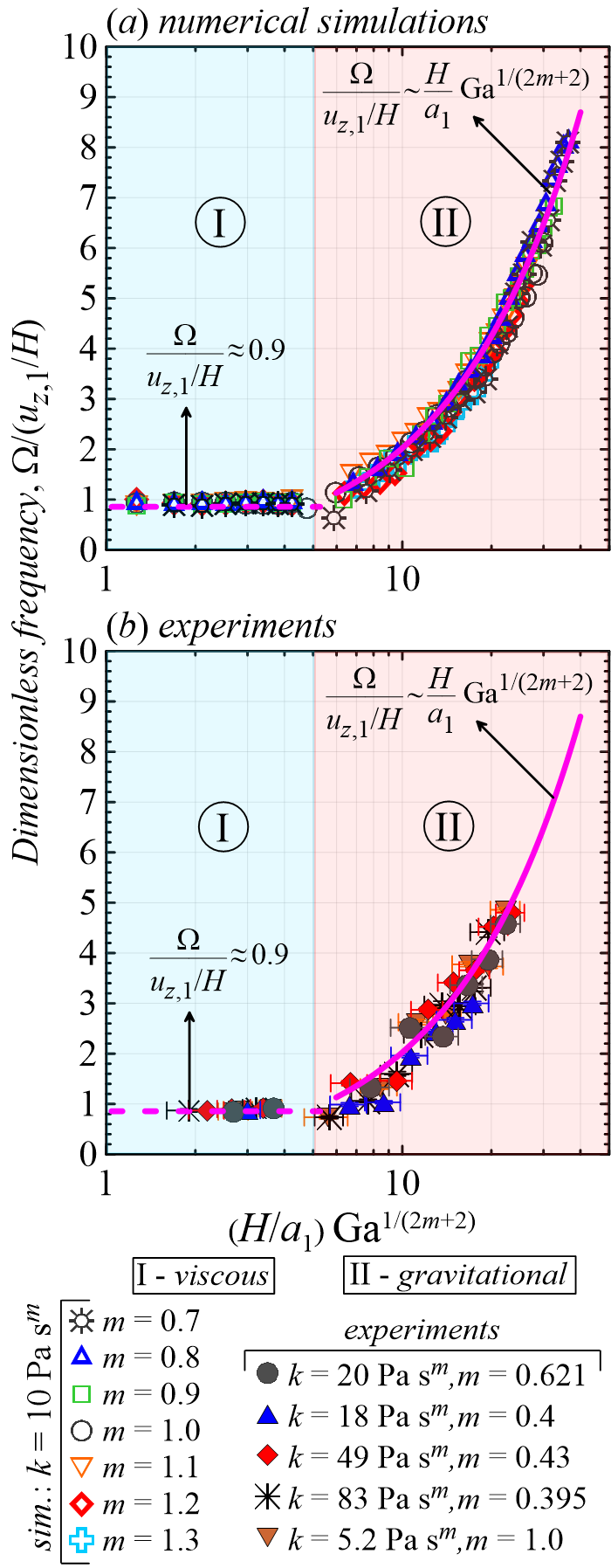}
\vspace*{-0.275cm}
\caption{$\Omega H/u_{z,1}$ as a function of $(H/a_1) \textrm{Ga}^{\frac{1}{2m+2}}$. Over 250 flow cases are displayed, the opened symbols in (\textit{a}) corresponding to numerical simulations, and the solid symbols in (\textit{b}) denoting experiments. The dashed line is given by Eq. \ref{eq: omega_fold_v} (viscous regime $\Rightarrow$ region I, in blue), while the solid one is given by Eq. \ref{eq: omega_fold_g2} (gravitational regime $\Rightarrow$ region II, in pink).}
\label{fig-9}
\end{figure}

\begin{figure*}%[h!]
\centering
\includegraphics[angle=0, scale=0.44]{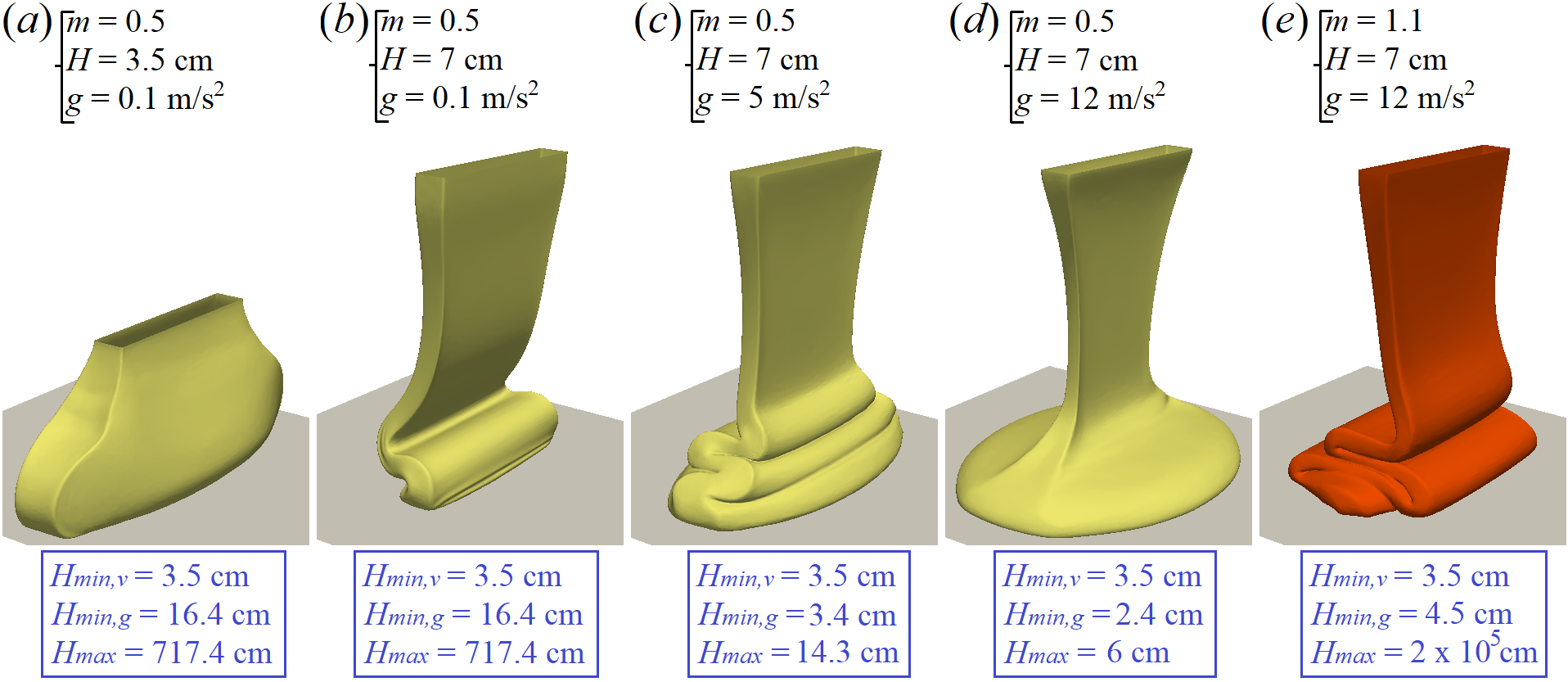}
\vspace*{-0.7cm}
\caption{Critical fall heights related to both the onset and cessation of folding for two power-law materials: $\rho = 1000$, kg/m$^3$, $k = 10$ Pa$\cdot$s$^m$, and $m = 0.5$ (shear-thinning; \textit{a}-\textit{d}; golden sheets); and $\rho = 1000$, kg/m$^3$, $k = 10$ Pa$\cdot$s$^m$, and $m = 1.1$ (shear-thickening; \textit{e}; orange sheet). For all analysed cases, the extrusion velocity $u_{z,0}$ is kept fixed at 0.1 m/s in the extrusion slot of dimension $a_0 = 0.7$ cm and $b_0 = 5.0$ cm. Their critical fall height values are stressed within the blue rectangles. Folding takes place when $H_{min,v} < H < H_{max}$ or $H_{min,g} < H < H_{max}$, where $H_{min,v}$, $H_{min,g}$, and $H_{max}$ are given by Eqs. \ref{eq: h_crit_min_v}, \ref{eq: h_crit_min_g}, and \ref{eq: h_max_2}, respectively.}
\label{fig-10}
\end{figure*}

Figure \ref{fig-9} shows $\Omega H/u_{z,1}$ as a function of $(H/a_1) \textrm{Ga}^{\frac{1}{2m+2}}$. The results are obtained from numerical simulations (Fig. \ref{fig-9}\textit{a}) and experiments (Fig. \ref{fig-9}\textit{b}; uncertainties are represented by the bars). They bring out the existence of both the viscous and the gravitational limit expressed by Eqs. \ref{eq: omega_fold_v}-\ref{eq: delta_fold_v}, and \ref{eq: omega_fold_g2}-\ref{eq: delta_fold_g2}, respectively, mapping out the transition between them. Indeed, these scaling laws are in good agreement with over 250 flow cases displayed in Fig. \ref{fig-9}. In the viscous regime, $\Omega/(u_{z,1}/H) \approx 0.9$ (as stressed by the magenta dashed line). It appears when $(H/a_1) \textrm{Ga}^{\frac{1}{2m+2}}$ is lower than a critical value of approximately 5, regardless the non-Newtonian nature of the sheets (blue region). From this critical value forward, bifurcation to the gravitational folding instabilities occurs (pink region) and, thus, the gravitational regime reaches its asymptotic form $\Omega/(u_{z,1}/H) = A (H/a_1) \textrm{Ga}^{\frac{1}{2m+2}}$, where $A \approx 0.2$, represented by the magenta solid line. Finally, it is important to observe that, for a Newtonian fluid, $m = 1$ and, consequently, $\Omega H / u_{z,1} \sim (H/a_1) \textrm{Ga}^{\frac{1}{4}}$ and $\delta/H \sim (a_1/H) \textrm{Ga}^{-\frac{1}{4}}$, as previously reported by Skorobogatiy and Mahadevan \cite{Skorobogatiy-00}, and Ribe \cite{Ribe-03}. In other words, our Newtonian results are rather in line with those presented by the referred authors (detailed comparisons are provided in the Supplemental Material). Nevertheless, their Newtonian folding equations cannot be used to predict shear-thinning/shear-thickening affected gravitational folding. These equations cannot be used to predict the folding suppression either, as discussed in the following lines.          
 
%%%%%%%%%%%%%%%%%%%%%%%%%%%%%%%%%%%%%%%%%%%%%%%%%%%%%%%%%%%%%%%%%%%%%%%%%%%%%%
\subsection{Folding suppression: onset and cessation} \label{FC}
%%%%%%%%%%%%%%%%%%%%%%%%%%%%%%%%%%%%%%%%%%%%%%%%%%%%%%%%%%%%%%%%%%%%%%%%%%%%%%

As indicated by Eqs. \ref{eq: omega_fold_v} and \ref{eq: delta_fold_v}, in the viscous regime, the energy related to folding deformations is proportional to $(u_{z,0}/H)(a_0/H)$, in contrast with the cost of compression that, in turn, is proportional to $u_{z,0}/H$. These energetic costs become comparable, however, when the sheet slenderness $a_0/H \sim 1$ and, consequently, $\delta \sim a_0 \sim H$. In other words, viscous folding vanishes when all the length scales of the problem ($\delta$, $a_0$, and $H$) collapse into a single one. For the viscous regime, such a collapse occurs when $H/a_0 \approx 5$ (this will be shown in details by the final diagrams of the present work), which leads to a critical viscous fall height
\begin{equation} 
H_{min,v} = B ~ a_0  \,  
\label{eq: h_crit_min_v} 
\end{equation}
below which viscous folding ceases (with $B \approx 5$).     

The rationale presented above can be equally applied to gravitational folding instabilities by assuming that $\delta \sim a_1 \sim b_1 \sim H$ in Eq. \ref{eq: delta_fold_g2} (and, consequently, $Q \sim a_1^2 u_{z,1}$). This leads to a critical gravitational fall height   
\begin{equation} 
H_{min,g} = C ~ \left( \frac{kQ^m}{\rho g} \right)^{\frac{1}{3m+1}} \,  
\label{eq: h_crit_min_g} 
\end{equation}
below which gravitational folding ceases (where $C \approx 3.2$ according to our numerical results, as will be shown in the following lines).

It is also worth noticing that, according to Eq. \ref{eq: delta_fold_g2}, $\delta$ is a decreasing function of buoyancy. Hence, when the sheet is highly exposed to gravity forces, one observes the formation of smaller folds. Eventually, $\delta$ decreases so much that it becomes comparable to $a_1$ ($\delta \sim a_1$) and, as a result, folding can no longer occur. From this point on, folding ceases and the sheet simply spreads. Rewriting the deduction of Eq. \ref{eq: delta_fold_g2} with $\delta \sim a_1$ (and considering that $\Delta u_{z} \sim u_{z,1}$), we find that, in the spreading scenario,  
\begin{equation} 
\frac{\rho u_{z,1}^2}{k(u_{z,1}/a_1)^m} \sim 1  ~~~ \textrm{(e.g. Reynolds number)} \, .  
\label{eq: h_max_1} 
\end{equation}
Assuming that, in this borderline scenario, the energy dissipation of the sheet is fully concentrated in the spreading region, and consequently $\Delta u_{z} \sim u_{z,1} \sim \sqrt{g H}$ (because the gravitational potential energy will be fully transferred into kinetic energy along the tail), we find a critical fall height    
\begin{equation} 
H_{max} = \frac{1}{g} \left(  \frac{D ~ k}{\rho ~ Q^{m/2}} \right)^{\frac{4}{4-3m}}   \,  
\label{eq: h_max_2} 
\end{equation}
above which folding ceases (where $D = 0.4^{2m-3}$, as will be shown in the final diagrams of this work, and $Q \sim a_1^2 u_{z,1}$ because in this particular flow scenario $a_1 \sim b_1$).   

Equations \ref{eq: h_crit_min_v}, \ref{eq: h_crit_min_g}, and \ref{eq: h_max_2} represent the power-law versions of those reported by Ribe et al. \cite{Ribe-12} by considering Newtonian fluids (see Eqs. 18, 19 and 20 of the referred work). These power-law-generalised predictions define, for each power-law sheet extruded at a flow rate $Q$ and gravity $g$, the fall height range in which folding occurs. Hence, when $H$ surmounts $H_{min,v}$ or $H_{min,g}$ without exceeding $H_{max}$, one recovers the viscous or the gravitational regime, respectively, of which finite-amplitude folding is highlighted in Subsection \ref{FR}. 

\begin{figure}%[h!]
\centering
\includegraphics[angle=0, scale=0.16]{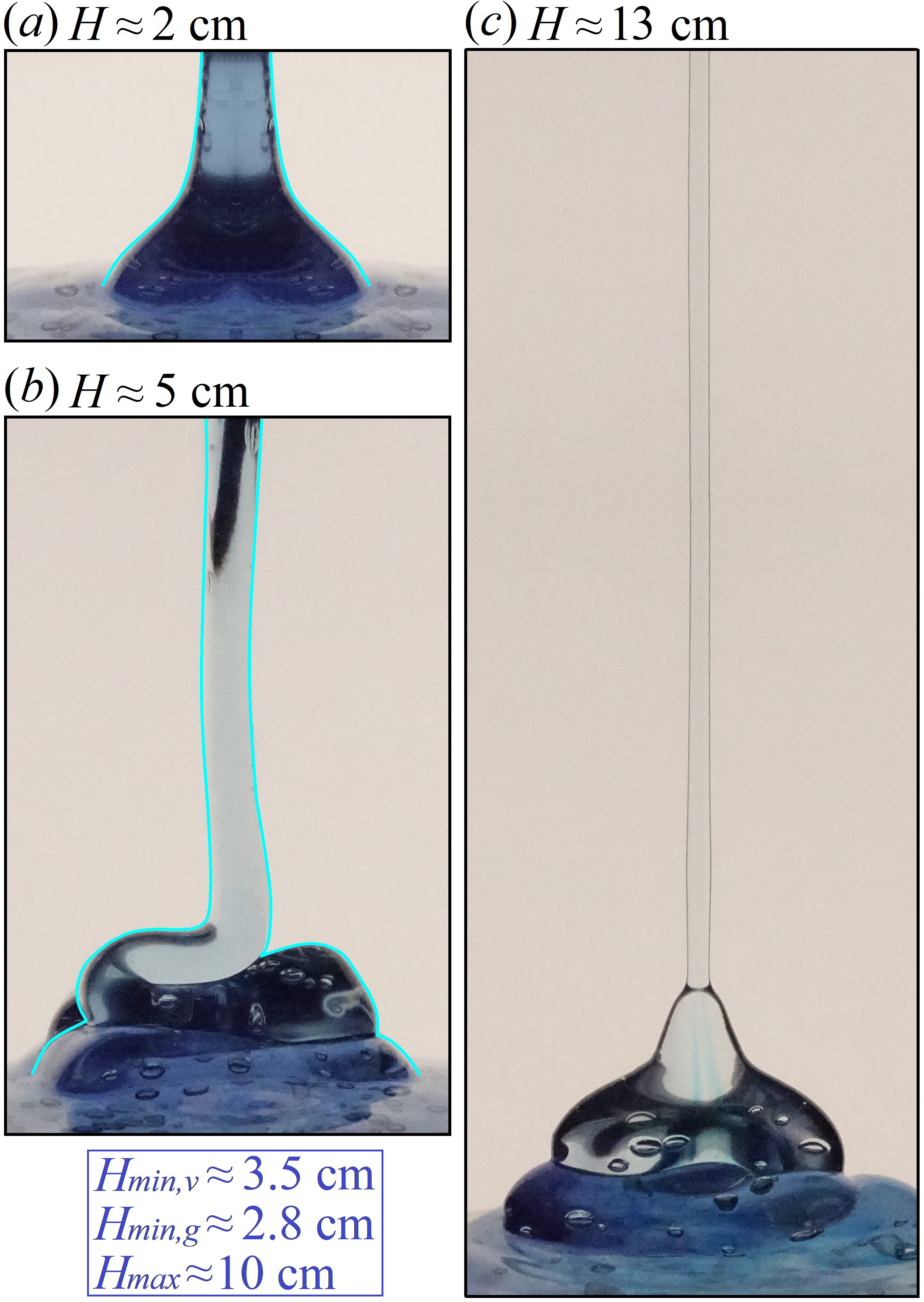}
\vspace*{-0.25cm}
\caption{Critical fall heights related to both the onset and cessation of folding are shown for a Carpobol suspension: $k = 18$ Pa$\cdot$s$^m$, and $m = 0.4$. The extrusion velocity $u_{z,0}$ is kept fixed at $\approx 0.1$ m/s in the extrusion slot of dimension $a_0 = 0.7$ cm and $b_0 = 5.0$ cm. At $H \approx 2$ cm (\textit{a}), the fluid simply spreads after hitting the substrate. Folding instabilities are observed for $2.8 \lessapprox  H \lessapprox 10$ cm, as illustrated in (\textit{b}; $H \approx 5$ cm); vanishing at higher $H$ (\textit{c}; $H \approx 13$ cm). These results are in agreement with the critical height fall values predicted by Eqs. \ref{eq: h_crit_min_v}, \ref{eq: h_crit_min_g} and \ref{eq: h_max_2}, and stressed within the blue rectangle.}
\label{fig-11}
\end{figure}

\begin{figure*}%[h!]
\centering
\includegraphics[angle=0, scale=0.32]{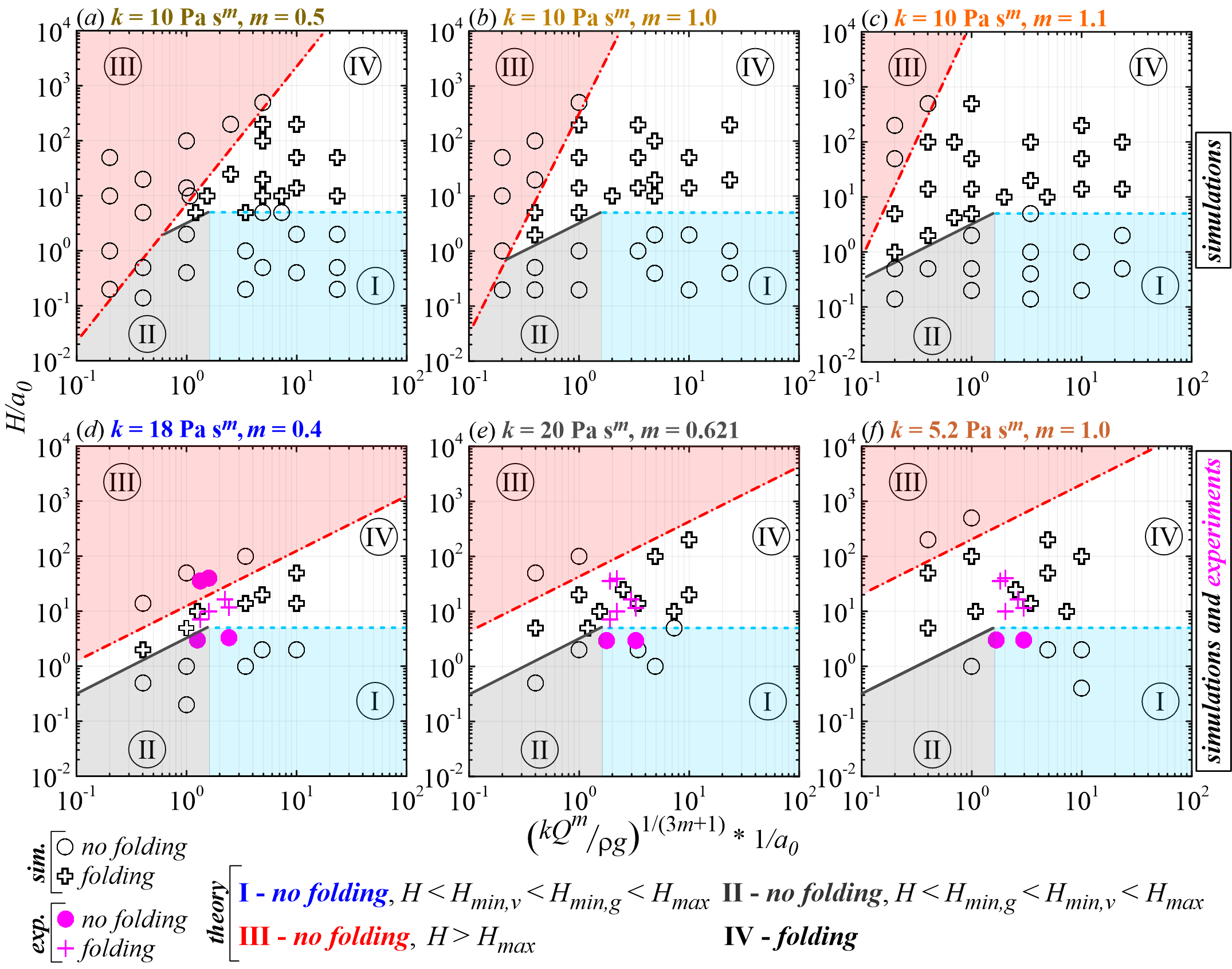}
\vspace*{-0.7cm}
\caption{Critical fall heights $H$ for the onset and cessation for steady folding are confronted with $\left( \frac{kQ^m}{\rho g} \right)^{\frac{1}{3m+1}}$. These quantities are made dimensionless by $a_0$. The opened symbols denote numerical simulations, while the closed ones indicate experiments, each symbol representing a flow case (159 numerical simulations and 21 experiments are considered; uncertainty bars are smaller than the symbols). The crosses represent flow cases for which folding occurs, while the circles indicate folding suppression. The top line figures (\textit{a}-\textit{c}) show exclusively numerical simulation results at $k = 10$ Pa$\cdot$s$^m$ for which three flow behaviour indexes are taken into account: $m = 0.5$ (Fig. \ref{fig-12}\textit{a}); $m = 1.0$ (\textit{b}); and $m = 1.1$ (\textit{c}). These top line diagrams are constructed by fixing $Q = 3.5$ x $10^{-5}$ m$^3$/s and $\rho = 1000$, kg/m$^3$, while $\mathcal{O} \left( 10^{-2} \right) \lesssim g \lesssim \mathcal{O} \left( 10^{3} \right)$ m/s$^2$ and $\mathcal{O} \left( 10^{-3} \right) \lesssim H \lesssim \mathcal{O} \left( 10^{0} \right)$ m. The bottom line figures (\textit{d}-\textit{f}) show both numerical (opened black symbols) and experimental (closed magenta symbols) results for three different fluids: (\textit{d}) $k = 18$ Pa$\cdot$s$^m$ and $m = 0.4$; (\textit{b}) $k = 20$ Pa$\cdot$s$^m$ and $m = 0.621$; (\textit{c}) $k = 5.2$ Pa$\cdot$s$^m$ and $m = 1.0$. These bottom line diagrams are constructed by fixing $Q = 3.5$ x $10^{-5}$ m$^3$/s, $\rho = 1000$, kg/m$^3$ and $g = 9.81 $ m/s$^2$, while $\mathcal{O} \left( 10^{-3} \right) \lesssim a_0 \lesssim \mathcal{O} \left( 10^{-2} \right)$ m and $\mathcal{O} \left( 10^{-3} \right) \lesssim H \lesssim \mathcal{O} \left( 10^{0} \right)$ m. For the six shown diagrams, four different regions are stressed. In the blue region (I), $H < H_{min,v} < H_{min,g} < H_{max}$. In the gray region (II), $H < H_{min,g} < H_{min,v} < H_{max}$. In the red region (III), $H > H_{max}$. As observed, folding occurs within the white region (IV) bounded by the critical heights defined by Eqs. \ref{eq: h_crit_min_v}, \ref{eq: h_crit_min_g}, and \ref{eq: h_max_2} and respectively denoted by the dashed blue line, the solid gray line, and the red dash-dotted line.}
\label{fig-12}
\end{figure*}

According to Eqs. \ref{eq: h_crit_min_g} and \ref{eq: h_max_2}, the fall height range in which folding takes place is highly affected not only by $g$, $Q$, and $\rho$, but also by $m$. The impact of some of these parameters on both the folding onset and cessation is explored in Fig. \ref{fig-10}, where different $H-g$ values are imposed to two power-law materials: $\rho = 1000$, kg/m$^3$, $k = 10$ Pa$\cdot$s$^m$, and $m = 0.5$ (shear-thinning; Fig. \ref{fig-10}\textit{a}-\textit{d}; golden sheets); and $\rho = 1000$, kg/m$^3$, $k = 10$ Pa$\cdot$s$^m$, and $m = 1.1$ (shear-thickening; Fig. \ref{fig-10}\textit{e}; orange sheet). For all analysed cases, the extrusion velocity $u_{z,0}$ is kept fixed at 0.1 m/s in the extrusion slot of dimension $a_0 = 0.7$ cm and $b_0 = 5.0$ cm. Their critical fall height values are stressed within the blue rectangles. At $g = 0.1$ m/s$^2$ and $H = 3.5$ cm, the shear-thinning fluid considered in Fig. \ref{fig-10}(\textit{a}) no folding occurs, since $H < H_{min,v}$. Consequently, the sheet only `inflates' after hitting the solid surface. When $H$ exceeds $H_{min,v}$, however, viscous-driven folding instabilities are trigged, as confirmed by the flow case illustrated in Fig. \ref{fig-10}(\textit{b}) for which $H = 7.0$ cm and $H_{min,v} = 3.5$ cm. Interestingly, the folding regime switch from viscous to gravitational by exposing the referred sheet to a more pronounced $g$ of 5 m/s$^2$ (Fig. \ref{fig-10}\textit{c}). Nevertheless, for $g > 11$ m/s$^2$, the critical height $H_{max}$ becomes smaller than $H = 7.0$ cm and, as a result, folding ceases (Fig. \ref{fig-10}\textit{d}). Lastly, it is important to observe that gravity-driven folding instabilities can be eventually recovered by simply increasing $m$, since shear-thickening sheets exhibit higher $H_{max}$, as shown in Fig. \ref{fig-10}(\textit{e}).      

The folding onset/cessation theoretical predictions are also confirmed by experimental results, as shown by Fig. \ref{fig-11}, in which the impact of the $k = 18$ Pa$\cdot$s$^m$ and $m = 0.4$ Carbopol suspension on its substrate is considered for three fall heights: $H \approx 2$ cm (Fig. \ref{fig-11}\textit{a}); $H \approx 5$ (Fig. \ref{fig-11}\textit{b}); and $H \approx 13$ cm (Fig. \ref{fig-11}\textit{c}). The extrusion velocity $u_{z,0}$ is kept fixed at $\approx 0.1$ m/s in the extrusion slot of dimension $a_0 = 0.7$ cm and $b_0 = 5.0$ cm. The critical fall height predicted by Eqs. \ref{eq: h_crit_min_v}, \ref{eq: h_crit_min_g}, and \ref{eq: h_max_2} are stressed within the blue rectangle. At $H \approx 2$ cm (Fig. \ref{fig-11}\textit{a}), the fluid simply spreads after hitting the substrate, because $ H < H_{min,v}$. At $H \approx 5$ cm, folding instabilities appear after the sheet impact, since $2.8 \lessapprox  H \lessapprox 10$ (Fig. \ref{fig-11}\textit{b}). However, once $H_{max} = 10$ cm is surpassed, folding ceases (Fig. \ref{fig-11}\textit{c}).     

The theoretical predictions expressed by Eqs. \ref{eq: h_crit_min_v}, \ref{eq: h_crit_min_g}, and \ref{eq: h_max_2} are systematically confirmed by Fig. \ref{fig-12}, where critical onset/cessation fall heights $H$ are confronted with $\left( \frac{kQ^m}{\rho g} \right)^{\frac{1}{3m+1}}$. These quantities are made dimensionless by $a_0$. The opened symbols denote numerical simulations, while the closed ones indicate experiments, each symbol representing a flow case (159 numerical simulations and 21 experiments are considered). The crosses represent flow cases for which folding occurs, while the circles indicate folding suppression. The top line figures show exclusively numerical simulation results at $k = 10$ Pa$\cdot$s$^m$ for which three flow behaviour indexes are taken into account: $m = 0.5$ (Fig. \ref{fig-12}\textit{a}); $m = 1.0$ (Fig. \ref{fig-12}\textit{b}); and $m = 1.1$ (Fig. \ref{fig-12}\textit{c}). These top line diagrams are constructed by fixing $Q = 3.5$ x $10^{-5}$ m$^3$/s and $\rho = 1000$, kg/m$^3$, while $\mathcal{O} \left( 10^{-2} \right) \lesssim g \lesssim \mathcal{O} \left( 10^{3} \right)$ m/s$^2$ and $\mathcal{O} \left( 10^{-3} \right) \lesssim H \lesssim \mathcal{O} \left( 10^{0} \right)$ m. The bottom line figures show both numerical (opened black symbols) and experimental (closed magenta symbols) results for three different fluids: (Fig. \ref{fig-12}\textit{d}) $k = 18$ Pa$\cdot$s$^m$ and $m = 0.4$; (Fig. \ref{fig-12}\textit{e}) $k = 20$ Pa$\cdot$s$^m$ and $m = 0.621$; (Fig. \ref{fig-12}\textit{f}) $k = 5.2$ Pa$\cdot$s$^m$ and $m = 1.0$. These bottom line diagrams are constructed by fixing $Q = 3.5$ x $10^{-5}$ m$^3$/s, $\rho = 1000$, kg/m$^3$ and $g = 9.81 $ m/s$^2$, while $\mathcal{O} \left( 10^{-3} \right) \lesssim a_0 \lesssim  \mathcal{O} \left( 10^{-2} \right)$ m and $\mathcal{O} \left( 10^{-3} \right) \lesssim H \lesssim \mathcal{O} \left( 10^{0} \right)$ m. The displayed diagrams are divided in four regions. Both the blue (I) and the gray (II) regions indicate fall heights $H$ smaller than the critical ones related to the folding onset. More specifically, in the blue region (I) $H < H_{min,v} < H_{min,g} < H_{max}$, while in the gray region (II) $H < H_{min,g} < H_{min,v} < H_{max}$. The red region (III), in turn, stresses fall heights that exceed $H_{max}$. Clearly, for both numerical and experimental results, folding takes place within the white region (IV) bounded by the critical heights defined by Eqs. \ref{eq: h_crit_min_v}, \ref{eq: h_crit_min_g}, and \ref{eq: h_max_2} and respectively denoted by the dashed blue line, the solid gray line, and the red dash-dotted line. Such a folding domain is highly affected by non-Newtonian aspects of the material, tending be suppressed by pronounced shear-thinning effects.     

%--------------------------------------------------------------------------------------------------------------------------------------------------------------------------------------
%--------------------------------------------------------------------------------------------------------------------------------------------------------------------------------------
\section{Concluding Remarks}
%--------------------------------------------------------------------------------------------------------------------------------------------------------------------------------------
%--------------------------------------------------------------------------------------------------------------------------------------------------------------------------------------
%\vspace*{-0.5cm}
We have stressed shear thickening (dilatant) and shear thinning (pseudoplastic) effects on the development of folding instabilities in non-Newtonian viscous sheets of which viscosity is given by a power-law constitutive equation. Our analyses were conducted through a mixed approach combining theoretical arguments (scaling laws), experiments and three dimensional numerical simulations. The numerical results were based on an adaptive variational multi-scale method for multiphase flows (power-law fluid and air), while Carpobol gel sheets were considered for the conducted experiments.

In short, two folding regimes are observed: (1) the viscous regimes; and (2) the gravitational one. However, only the latter is affected rheological aspects of the material and, consequently, by shear thinning/thickening manifestations. In this folding regime, the sheet is highly exposed to gravity and, thus, the folding dynamics is driven by balanced by gravity and viscous forces (gravitational regime), both the folding frequency, $\Omega$, and the folding amplitude, $\delta$ (made dimensionless by $u_{z,1}/H$ and $H$, respectively) appear as a function of the sheet slenderness, the Galileo number $\textrm{Ga}$, and the consistency index $m$. More specifically, $\Omega H / u_{z,1} \sim (H/a_1) \textrm{Ga}^{\frac{1}{2m+2}}$, while $\delta/H \sim (a_1/H) \textrm{Ga}^{\frac{-1}{2m+2}}$. For instance, for a Newtonian fluid, $m = 1$ and, consequently, $\Omega H / u_{z,1} \sim (H/a_1) \textrm{Ga}^{\frac{1}{4}}$ and  $\delta/H \sim (a_1/H) \textrm{Ga}^{-\frac{1}{4}}$, as previously reported by Skorobogatiy and Mahadevan \cite{Skorobogatiy-00}, and Ribe \cite{Ribe-03}. In addition, highly shear thickening materials develop large amplitude and low frequency instabilities, which, in contrast, tend to be suppressed by shear thinning effects, and eventually cease. 

Both the shear thinning and the shear thickening effects on the folding onset/cessation were carefully analysed as well. The new non-Newtonian folding onset/cessation criteria presented here are in good agreement with the obtained experimental/numerical results.           

Finally, since complex materials are highly diffused in industrial domains, it would be interesting to consider in future works supplemental non-Newtonian effects on the folding of sheets, such as those related to the yield stress and/or elasticity.  
%\vspace{2cm}

%%%%%%%%%%%%%%%%%%%%%%%%%%%%%%%%%%%%%%%%%%%%%%%%%%%%%%%%%%%%%%%%%%%%%%%%%%%%%%
%%%%%%%%%%%%%%%%%%%%%%%%%%%%%%%%%%%%%%%%%%%%%%%%%%%%%%%%%%%%%%%%%%%%%%%%%%%%%%
\textit{\textbf{Acknowledgements:}} 
%%%%%%%%%%%%%%%%%%%%%%%%%%%%%%%%%%%%%%%%%%%%%%%%%%%%%%%%%%%%%%%%%%%%%%%%%%%%%%
%%%%%%%%%%%%%%%%%%%%%%%%%%%%%%%%%%%%%%%%%%%%%%%%%%%%%%%%%%%%%%%%%%%%%%%%%%%%%%
The authors would like to thank Mrs. Camila Borgo for her great help with the experiments performed here, as well as Dr. Romain Castellani (PSL Research University, MINES ParisTech, CEMEF) for the rheological measurements displayed in Fig. \ref{fig-3}. The authors also would like to thank the PSL Research University for its support under the program `Investissements d'Avenir' launched by the French Government and implemented by the French National Research Agency (ANR) with the reference ANR-10-IDEX-0001-02 PSL.

%\newpage
%\clearpage

%%%%%%%%%%%%%%%%%%%%%%%%%%%%%%%%%%%%%%%%%%%%%%%%%%%%%%%%%%%%%%%%%%%%%%%%%%%%%%
%%%%%%%%%%%%%%%%%%%%%%%%%%%%%%%%%%%%%%%%%%%%%%%%%%%%%%%%%%%%%%%%%%%%%%%%%%%%%%
\bibliographystyle{apsrev4-2}
\bibliography{Buckling-ref}% Produces the bibliography via BibTeX.

%apsrev4-2.bst 2019-01-14 (MD) hand-edited version of apsrev4-1.bst
%Control: key (0)
%Control: author (72) initials jnrlst
%Control: editor formatted (1) identically to author
%Control: production of article title (-1) disabled
%Control: page (0) single
%Control: year (1) truncated
%Control: production of eprint (0) enabled
\begin{thebibliography}{35}%
\makeatletter
\providecommand \@ifxundefined [1]{%
 \@ifx{#1\undefined}
}%
\providecommand \@ifnum [1]{%
 \ifnum #1\expandafter \@firstoftwo
 \else \expandafter \@secondoftwo
 \fi
}%
\providecommand \@ifx [1]{%
 \ifx #1\expandafter \@firstoftwo
 \else \expandafter \@secondoftwo
 \fi
}%
\providecommand \natexlab [1]{#1}%
\providecommand \enquote  [1]{``#1''}%
\providecommand \bibnamefont  [1]{#1}%
\providecommand \bibfnamefont [1]{#1}%
\providecommand \citenamefont [1]{#1}%
\providecommand \href@noop [0]{\@secondoftwo}%
\providecommand \href [0]{\begingroup \@sanitize@url \@href}%
\providecommand \@href[1]{\@@startlink{#1}\@@href}%
\providecommand \@@href[1]{\endgroup#1\@@endlink}%
\providecommand \@sanitize@url [0]{\catcode `\\12\catcode `\$12\catcode
  `\&12\catcode `\#12\catcode `\^12\catcode `\_12\catcode `\%12\relax}%
\providecommand \@@startlink[1]{}%
\providecommand \@@endlink[0]{}%
\providecommand \url  [0]{\begingroup\@sanitize@url \@url }%
\providecommand \@url [1]{\endgroup\@href {#1}{\urlprefix }}%
\providecommand \urlprefix  [0]{URL }%
\providecommand \Eprint [0]{\href }%
\providecommand \doibase [0]{https://doi.org/}%
\providecommand \selectlanguage [0]{\@gobble}%
\providecommand \bibinfo  [0]{\@secondoftwo}%
\providecommand \bibfield  [0]{\@secondoftwo}%
\providecommand \translation [1]{[#1]}%
\providecommand \BibitemOpen [0]{}%
\providecommand \bibitemStop [0]{}%
\providecommand \bibitemNoStop [0]{.\EOS\space}%
\providecommand \EOS [0]{\spacefactor3000\relax}%
\providecommand \BibitemShut  [1]{\csname bibitem#1\endcsname}%
\let\auto@bib@innerbib\@empty
%</preamble>
\bibitem [{\citenamefont {Barnes}\ and\ \citenamefont
  {Woodcock}(1958)}]{Barnes-58}%
  \BibitemOpen
  \bibfield  {author} {\bibinfo {author} {\bibfnamefont {G.}~\bibnamefont
  {Barnes}}\ and\ \bibinfo {author} {\bibfnamefont {R.}~\bibnamefont
  {Woodcock}},\ }\href@noop {} {\bibfield  {journal} {\bibinfo  {journal}
  {American Journal of Physics}\ }\textbf {\bibinfo {volume} {26}},\ \bibinfo
  {pages} {205} (\bibinfo {year} {1958})}\BibitemShut {NoStop}%
\bibitem [{\citenamefont {Taylor}(1969)}]{Taylor-69}%
  \BibitemOpen
  \bibfield  {author} {\bibinfo {author} {\bibfnamefont {G.~I.}\ \bibnamefont
  {Taylor}},\ }\href@noop {} {\bibfield  {journal} {\bibinfo  {journal}
  {Proceedings of the Twelfth International Congress of Applied Mechanics,
  Stanford, 1968}\ }\textbf {\bibinfo {volume} {Springer-Verlag, Berlin}},\
  \bibinfo {pages} {382} (\bibinfo {year} {1969})}\BibitemShut {NoStop}%
\bibitem [{\citenamefont {Cruickshank}(1988)}]{Cruickshank-88}%
  \BibitemOpen
  \bibfield  {author} {\bibinfo {author} {\bibfnamefont {J.~O.}\ \bibnamefont
  {Cruickshank}},\ }\href@noop {} {\bibfield  {journal} {\bibinfo  {journal}
  {Journal of Fluid Mechanics}\ }\textbf {\bibinfo {volume} {193}},\ \bibinfo
  {pages} {111} (\bibinfo {year} {1988})}\BibitemShut {NoStop}%
\bibitem [{\citenamefont {Yarin}\ and\ \citenamefont
  {Tchavdarov}(1996)}]{Yarin-96}%
  \BibitemOpen
  \bibfield  {author} {\bibinfo {author} {\bibfnamefont {A.~L.}\ \bibnamefont
  {Yarin}}\ and\ \bibinfo {author} {\bibfnamefont {B.~M.}\ \bibnamefont
  {Tchavdarov}},\ }\href@noop {} {\bibfield  {journal} {\bibinfo  {journal}
  {Journal of Fluid Mechanics}\ }\textbf {\bibinfo {volume} {307}},\ \bibinfo
  {pages} {85} (\bibinfo {year} {1996})}\BibitemShut {NoStop}%
\bibitem [{\citenamefont {Mahadevan}\ \emph {et~al.}(1998)\citenamefont
  {Mahadevan}, \citenamefont {Ryu},\ and\ \citenamefont
  {Samuel}}]{Mahadevan-98}%
  \BibitemOpen
  \bibfield  {author} {\bibinfo {author} {\bibfnamefont {L.}~\bibnamefont
  {Mahadevan}}, \bibinfo {author} {\bibfnamefont {W.~S.}\ \bibnamefont {Ryu}},\
  and\ \bibinfo {author} {\bibfnamefont {A.~D.~T.}\ \bibnamefont {Samuel}},\
  }\href@noop {} {\bibfield  {journal} {\bibinfo  {journal} {Nature}\ }\textbf
  {\bibinfo {volume} {392}},\ \bibinfo {pages} {140} (\bibinfo {year}
  {1998})}\BibitemShut {NoStop}%
\bibitem [{\citenamefont {Mahadevan}\ \emph {et~al.}(2000)\citenamefont
  {Mahadevan}, \citenamefont {Ryu},\ and\ \citenamefont
  {Samuel}}]{Mahadevan-00}%
  \BibitemOpen
  \bibfield  {author} {\bibinfo {author} {\bibfnamefont {L.}~\bibnamefont
  {Mahadevan}}, \bibinfo {author} {\bibfnamefont {W.~S.}\ \bibnamefont {Ryu}},\
  and\ \bibinfo {author} {\bibfnamefont {A.~D.~T.}\ \bibnamefont {Samuel}},\
  }\href@noop {} {\bibfield  {journal} {\bibinfo  {journal} {Nature}\ }\textbf
  {\bibinfo {volume} {403}},\ \bibinfo {pages} {502} (\bibinfo {year}
  {2000})}\BibitemShut {NoStop}%
\bibitem [{\citenamefont {{Le Merrer}}\ \emph {et~al.}(2012)\citenamefont {{Le
  Merrer}}, \citenamefont {Qu\'er\'e},\ and\ \citenamefont
  {Clanet}}]{Merrer-12}%
  \BibitemOpen
  \bibfield  {author} {\bibinfo {author} {\bibfnamefont {M.}~\bibnamefont {{Le
  Merrer}}}, \bibinfo {author} {\bibfnamefont {D.}~\bibnamefont {Qu\'er\'e}},\
  and\ \bibinfo {author} {\bibfnamefont {C.}~\bibnamefont {Clanet}},\
  }\href@noop {} {\bibfield  {journal} {\bibinfo  {journal} {Physical Review
  Letters}\ }\textbf {\bibinfo {volume} {109}},\ \bibinfo {pages} {064502}
  (\bibinfo {year} {2012})}\BibitemShut {NoStop}%
\bibitem [{\citenamefont {Ribe}(2003)}]{Ribe-03}%
  \BibitemOpen
  \bibfield  {author} {\bibinfo {author} {\bibfnamefont {N.~M.}\ \bibnamefont
  {Ribe}},\ }\href@noop {} {\bibfield  {journal} {\bibinfo  {journal} {Physical
  Review E}\ }\textbf {\bibinfo {volume} {68}},\ \bibinfo {pages} {036305}
  (\bibinfo {year} {2003})}\BibitemShut {NoStop}%
\bibitem [{\citenamefont {Ribe}\ \emph {et~al.}(2006)\citenamefont {Ribe},
  \citenamefont {Huppert}, \citenamefont {Hallworth}, \citenamefont {Habibi},\
  and\ \citenamefont {Bonn}}]{Ribe-06}%
  \BibitemOpen
  \bibfield  {author} {\bibinfo {author} {\bibfnamefont {N.~M.}\ \bibnamefont
  {Ribe}}, \bibinfo {author} {\bibfnamefont {H.~E.}\ \bibnamefont {Huppert}},
  \bibinfo {author} {\bibfnamefont {M.~A.}\ \bibnamefont {Hallworth}}, \bibinfo
  {author} {\bibfnamefont {M.}~\bibnamefont {Habibi}},\ and\ \bibinfo {author}
  {\bibfnamefont {D.}~\bibnamefont {Bonn}},\ }\href@noop {} {\bibfield
  {journal} {\bibinfo  {journal} {Journal of Fluid Mechanics}\ }\textbf
  {\bibinfo {volume} {555}},\ \bibinfo {pages} {275} (\bibinfo {year}
  {2006})}\BibitemShut {NoStop}%
\bibitem [{\citenamefont {Ribe}\ \emph {et~al.}(2012)\citenamefont {Ribe},
  \citenamefont {Habibi},\ and\ \citenamefont {Bonn}}]{Ribe-12}%
  \BibitemOpen
  \bibfield  {author} {\bibinfo {author} {\bibfnamefont {N.~M.}\ \bibnamefont
  {Ribe}}, \bibinfo {author} {\bibfnamefont {M.}~\bibnamefont {Habibi}},\ and\
  \bibinfo {author} {\bibfnamefont {D.}~\bibnamefont {Bonn}},\ }\href@noop {}
  {\bibfield  {journal} {\bibinfo  {journal} {Annual Review of Fluid
  Mechanics}\ }\textbf {\bibinfo {volume} {44}},\ \bibinfo {pages} {249}
  (\bibinfo {year} {2012})}\BibitemShut {NoStop}%
\bibitem [{\citenamefont {Tian}\ \emph {et~al.}(2020)\citenamefont {Tian},
  \citenamefont {Ribe}, \citenamefont {Wu},\ and\ \citenamefont
  {Shum}}]{Tian-20}%
  \BibitemOpen
  \bibfield  {author} {\bibinfo {author} {\bibfnamefont {J.}~\bibnamefont
  {Tian}}, \bibinfo {author} {\bibfnamefont {N.}~\bibnamefont {Ribe}}, \bibinfo
  {author} {\bibfnamefont {X.}~\bibnamefont {Wu}},\ and\ \bibinfo {author}
  {\bibfnamefont {H.~C.}\ \bibnamefont {Shum}},\ }\href@noop {} {\bibfield
  {journal} {\bibinfo  {journal} {Physical Review Letters}\ }\textbf {\bibinfo
  {volume} {125}},\ \bibinfo {pages} {104502} (\bibinfo {year}
  {2020})}\BibitemShut {NoStop}%
\bibitem [{\citenamefont {Pilkington}(1969)}]{Pilkington-69}%
  \BibitemOpen
  \bibfield  {author} {\bibinfo {author} {\bibfnamefont {L.~A.~B.}\
  \bibnamefont {Pilkington}},\ }\href@noop {} {\bibfield  {journal} {\bibinfo
  {journal} {Transport Phenomena and Fluid Mechanics, AIChE Journal}\ }\textbf
  {\bibinfo {volume} {314}},\ \bibinfo {pages} {1} (\bibinfo {year}
  {1969})}\BibitemShut {NoStop}%
\bibitem [{\citenamefont {Pearson}(1985)}]{Pearson-85}%
  \BibitemOpen
  \bibfield  {author} {\bibinfo {author} {\bibfnamefont {J.}~\bibnamefont
  {Pearson}},\ }\href@noop {} {\bibfield  {journal} {\bibinfo  {journal}
  {Elsevier, Amsterdam}\ } (\bibinfo {year} {1985})}\BibitemShut {NoStop}%
\bibitem [{\citenamefont {Griffiths}\ and\ \citenamefont
  {Turner}(1988)}]{Griffiths-88}%
  \BibitemOpen
  \bibfield  {author} {\bibinfo {author} {\bibfnamefont {R.~W.}\ \bibnamefont
  {Griffiths}}\ and\ \bibinfo {author} {\bibfnamefont {J.~S.}\ \bibnamefont
  {Turner}},\ }\href@noop {} {\bibfield  {journal} {\bibinfo  {journal}
  {Geophysical Journal International}\ }\textbf {\bibinfo {volume} {95}},\
  \bibinfo {pages} {397} (\bibinfo {year} {1988})}\BibitemShut {NoStop}%
\bibitem [{\citenamefont {Johnson}\ and\ \citenamefont
  {Fletcher}(1994)}]{Johnson-94}%
  \BibitemOpen
  \bibfield  {author} {\bibinfo {author} {\bibfnamefont {A.~M.}\ \bibnamefont
  {Johnson}}\ and\ \bibinfo {author} {\bibfnamefont {R.~C.}\ \bibnamefont
  {Fletcher}},\ }\href@noop {} {\bibfield  {journal} {\bibinfo  {journal}
  {Columbia University, New York}\ } (\bibinfo {year} {1994})}\BibitemShut
  {NoStop}%
\bibitem [{\citenamefont {Rasschaert}\ \emph {et~al.}(2018)\citenamefont
  {Rasschaert}, \citenamefont {Talansier}, \citenamefont {Bl\'es\`es},
  \citenamefont {Magnin},\ and\ \citenamefont {Lambert}}]{Rasschaert-18}%
  \BibitemOpen
  \bibfield  {author} {\bibinfo {author} {\bibfnamefont {F.}~\bibnamefont
  {Rasschaert}}, \bibinfo {author} {\bibfnamefont {E.}~\bibnamefont
  {Talansier}}, \bibinfo {author} {\bibfnamefont {D.}~\bibnamefont
  {Bl\'es\`es}}, \bibinfo {author} {\bibfnamefont {A.}~\bibnamefont {Magnin}},\
  and\ \bibinfo {author} {\bibfnamefont {M.}~\bibnamefont {Lambert}},\
  }\href@noop {} {\bibfield  {journal} {\bibinfo  {journal} {Transport
  Phenomena and Fluid Mechanics, AIChE Journal}\ }\textbf {\bibinfo {volume}
  {64}},\ \bibinfo {pages} {1117} (\bibinfo {year} {2018})}\BibitemShut
  {NoStop}%
\bibitem [{\citenamefont {Habibi}\ \emph {et~al.}(2014)\citenamefont {Habibi},
  \citenamefont {Hosseini}, \citenamefont {Khatami},\ and\ \citenamefont
  {Ribe}}]{Habibi-14}%
  \BibitemOpen
  \bibfield  {author} {\bibinfo {author} {\bibfnamefont {M.}~\bibnamefont
  {Habibi}}, \bibinfo {author} {\bibfnamefont {S.~H.}\ \bibnamefont
  {Hosseini}}, \bibinfo {author} {\bibfnamefont {M.~H.}\ \bibnamefont
  {Khatami}},\ and\ \bibinfo {author} {\bibfnamefont {N.~M.}\ \bibnamefont
  {Ribe}},\ }\href@noop {} {\bibfield  {journal} {\bibinfo  {journal} {Physics
  of Fluids}\ }\textbf {\bibinfo {volume} {26}},\ \bibinfo {pages} {024101}
  (\bibinfo {year} {2014})}\BibitemShut {NoStop}%
\bibitem [{\citenamefont {Ribe}(2017)}]{Ribe-17}%
  \BibitemOpen
  \bibfield  {author} {\bibinfo {author} {\bibfnamefont {N.~M.}\ \bibnamefont
  {Ribe}},\ }\href@noop {} {\bibfield  {journal} {\bibinfo  {journal} {Journal
  of Fluid Mechanics}\ }\textbf {\bibinfo {volume} {812}},\ \bibinfo {pages}
  {R2} (\bibinfo {year} {2017})}\BibitemShut {NoStop}%
\bibitem [{\citenamefont {Tom\'e}\ \emph {et~al.}(2019)\citenamefont {Tom\'e},
  \citenamefont {Araujo}, \citenamefont {Evans},\ and\ \citenamefont
  {Mckee}}]{Tome-19}%
  \BibitemOpen
  \bibfield  {author} {\bibinfo {author} {\bibfnamefont {M.~F.}\ \bibnamefont
  {Tom\'e}}, \bibinfo {author} {\bibfnamefont {M.~T.}\ \bibnamefont {Araujo}},
  \bibinfo {author} {\bibfnamefont {J.}~\bibnamefont {Evans}},\ and\ \bibinfo
  {author} {\bibfnamefont {S.}~\bibnamefont {Mckee}},\ }\href@noop {}
  {\bibfield  {journal} {\bibinfo  {journal} {Journal of Non-Newtonian Fluid
  Mechanics}\ }\textbf {\bibinfo {volume} {263}},\ \bibinfo {pages} {104}
  (\bibinfo {year} {2019})}\BibitemShut {NoStop}%
\bibitem [{\citenamefont {Pereira}\ \emph {et~al.}(2019)\citenamefont
  {Pereira}, \citenamefont {Larcher}, \citenamefont {Hachem},\ and\
  \citenamefont {Valette}}]{Pereira-19b}%
  \BibitemOpen
  \bibfield  {author} {\bibinfo {author} {\bibfnamefont {A.}~\bibnamefont
  {Pereira}}, \bibinfo {author} {\bibfnamefont {A.}~\bibnamefont {Larcher}},
  \bibinfo {author} {\bibfnamefont {E.}~\bibnamefont {Hachem}},\ and\ \bibinfo
  {author} {\bibfnamefont {R.}~\bibnamefont {Valette}},\ }\href@noop {}
  {\bibfield  {journal} {\bibinfo  {journal} {Computers and Fluids}\ }\textbf
  {\bibinfo {volume} {190}},\ \bibinfo {pages} {514} (\bibinfo {year}
  {2019})}\BibitemShut {NoStop}%
\bibitem [{\citenamefont {Ostwald}(1925)}]{Ostwald-25}%
  \BibitemOpen
  \bibfield  {author} {\bibinfo {author} {\bibfnamefont {W.}~\bibnamefont
  {Ostwald}},\ }\href@noop {} {\bibfield  {journal} {\bibinfo  {journal}
  {Kolloid-Z}\ }\textbf {\bibinfo {volume} {36}},\ \bibinfo {pages} {99–117}
  (\bibinfo {year} {1925})}\BibitemShut {NoStop}%
\bibitem [{\citenamefont {Bird}\ \emph {et~al.}(1987)\citenamefont {Bird},
  \citenamefont {Armstrong},\ and\ \citenamefont {Hassager}}]{Bird-87}%
  \BibitemOpen
  \bibfield  {author} {\bibinfo {author} {\bibfnamefont {R.~B.}\ \bibnamefont
  {Bird}}, \bibinfo {author} {\bibfnamefont {R.~C.}\ \bibnamefont
  {Armstrong}},\ and\ \bibinfo {author} {\bibfnamefont {O.}~\bibnamefont
  {Hassager}},\ }\href@noop {} {\bibfield  {journal} {\bibinfo  {journal}
  {Wiley-Interscience, New York}\ }\textbf {\bibinfo {volume} {2nd edition}},\
  \bibinfo {pages} {172} (\bibinfo {year} {1987})}\BibitemShut {NoStop}%
\bibitem [{\citenamefont {Skorobogatiy}\ and\ \citenamefont
  {Mahadevan}(2000)}]{Skorobogatiy-00}%
  \BibitemOpen
  \bibfield  {author} {\bibinfo {author} {\bibfnamefont {M.}~\bibnamefont
  {Skorobogatiy}}\ and\ \bibinfo {author} {\bibfnamefont {L.}~\bibnamefont
  {Mahadevan}},\ }\href@noop {} {\bibfield  {journal} {\bibinfo  {journal}
  {Europhysics Letters}\ }\textbf {\bibinfo {volume} {52}},\ \bibinfo {pages}
  {532} (\bibinfo {year} {2000})}\BibitemShut {NoStop}%
\bibitem [{\citenamefont {Coupez}\ and\ \citenamefont
  {Hachem}(2013)}]{Coupez-13}%
  \BibitemOpen
  \bibfield  {author} {\bibinfo {author} {\bibfnamefont {T.}~\bibnamefont
  {Coupez}}\ and\ \bibinfo {author} {\bibfnamefont {E.}~\bibnamefont
  {Hachem}},\ }\href@noop {} {\bibfield  {journal} {\bibinfo  {journal}
  {Computer Methods in Applied Mechanics and Engineering}\ }\textbf {\bibinfo
  {volume} {267}},\ \bibinfo {pages} {65} (\bibinfo {year} {2013})}\BibitemShut
  {NoStop}%
\bibitem [{\citenamefont {Valette}\ \emph {et~al.}(2019)\citenamefont
  {Valette}, \citenamefont {Hachem}, \citenamefont {Khalloufi}, \citenamefont
  {Pereira}, \citenamefont {Mackley},\ and\ \citenamefont
  {Butler}}]{Valette-19}%
  \BibitemOpen
  \bibfield  {author} {\bibinfo {author} {\bibfnamefont {R.}~\bibnamefont
  {Valette}}, \bibinfo {author} {\bibfnamefont {E.}~\bibnamefont {Hachem}},
  \bibinfo {author} {\bibfnamefont {M.}~\bibnamefont {Khalloufi}}, \bibinfo
  {author} {\bibfnamefont {A.~S.}\ \bibnamefont {Pereira}}, \bibinfo {author}
  {\bibfnamefont {M.~R.}\ \bibnamefont {Mackley}},\ and\ \bibinfo {author}
  {\bibfnamefont {S.~A.}\ \bibnamefont {Butler}},\ }\href@noop {} {\bibfield
  {journal} {\bibinfo  {journal} {Journal of Non-Newtonian Fluid Mechanics}\
  }\textbf {\bibinfo {volume} {263}},\ \bibinfo {pages} {130} (\bibinfo {year}
  {2019})}\BibitemShut {NoStop}%
\bibitem [{\citenamefont {Valette}\ \emph {et~al.}(2021)\citenamefont
  {Valette}, \citenamefont {Pereira}, \citenamefont {Riber}, \citenamefont
  {Sardo}, \citenamefont {Larcher},\ and\ \citenamefont {Hachem}}]{Valette-20}%
  \BibitemOpen
  \bibfield  {author} {\bibinfo {author} {\bibfnamefont {V.}~\bibnamefont
  {Valette}}, \bibinfo {author} {\bibfnamefont {A.}~\bibnamefont {Pereira}},
  \bibinfo {author} {\bibfnamefont {S.}~\bibnamefont {Riber}}, \bibinfo
  {author} {\bibfnamefont {L.}~\bibnamefont {Sardo}}, \bibinfo {author}
  {\bibfnamefont {A.}~\bibnamefont {Larcher}},\ and\ \bibinfo {author}
  {\bibfnamefont {E.}~\bibnamefont {Hachem}},\ }\href
  {https://doi.org/https://doi.org/10.1016/j.jnnfm.2020.104447} {\bibfield
  {journal} {\bibinfo  {journal} {Journal of Non-Newtonian Fluid Mechanics}\
  }\textbf {\bibinfo {volume} {287}},\ \bibinfo {pages} {104447} (\bibinfo
  {year} {2021})}\BibitemShut {NoStop}%
\bibitem [{\citenamefont {Pereira}\ \emph {et~al.}(2020)\citenamefont
  {Pereira}, \citenamefont {Hachem},\ and\ \citenamefont
  {Valette}}]{Pereira-20}%
  \BibitemOpen
  \bibfield  {author} {\bibinfo {author} {\bibfnamefont {A.}~\bibnamefont
  {Pereira}}, \bibinfo {author} {\bibfnamefont {E.}~\bibnamefont {Hachem}},\
  and\ \bibinfo {author} {\bibfnamefont {R.}~\bibnamefont {Valette}},\
  }\href@noop {} {\bibfield  {journal} {\bibinfo  {journal} {Journal of
  Non-Newtonian Fluid Mechanics}\ }\textbf {\bibinfo {volume} {282}},\ \bibinfo
  {pages} {104321} (\bibinfo {year} {2020})}\BibitemShut {NoStop}%
\bibitem [{\citenamefont {Papanastasiou}(1987)}]{Papanastasiou-87}%
  \BibitemOpen
  \bibfield  {author} {\bibinfo {author} {\bibfnamefont {T.}~\bibnamefont
  {Papanastasiou}},\ }\href@noop {} {\bibfield  {journal} {\bibinfo  {journal}
  {Journal of Rheology}\ }\textbf {\bibinfo {volume} {31}},\ \bibinfo {pages}
  {385–404} (\bibinfo {year} {1987})}\BibitemShut {NoStop}%
\bibitem [{\citenamefont {Riber}\ \emph {et~al.}(2016)\citenamefont {Riber},
  \citenamefont {Valette}, \citenamefont {Mesri},\ and\ \citenamefont
  {Hachem}}]{Riber-16}%
  \BibitemOpen
  \bibfield  {author} {\bibinfo {author} {\bibfnamefont {S.}~\bibnamefont
  {Riber}}, \bibinfo {author} {\bibfnamefont {R.}~\bibnamefont {Valette}},
  \bibinfo {author} {\bibfnamefont {Y.}~\bibnamefont {Mesri}},\ and\ \bibinfo
  {author} {\bibfnamefont {E.}~\bibnamefont {Hachem}},\ }\href@noop {}
  {\bibfield  {journal} {\bibinfo  {journal} {Computers and Fluids}\ }\textbf
  {\bibinfo {volume} {138}},\ \bibinfo {pages} {51} (\bibinfo {year}
  {2016})}\BibitemShut {NoStop}%
\bibitem [{\citenamefont {Hachem}\ \emph {et~al.}(2016)\citenamefont {Hachem},
  \citenamefont {Khalloufi}, \citenamefont {Bruchon}, \citenamefont {Valette},\
  and\ \citenamefont {Mesri}}]{Hachem-16}%
  \BibitemOpen
  \bibfield  {author} {\bibinfo {author} {\bibfnamefont {E.}~\bibnamefont
  {Hachem}}, \bibinfo {author} {\bibfnamefont {M.}~\bibnamefont {Khalloufi}},
  \bibinfo {author} {\bibfnamefont {J.}~\bibnamefont {Bruchon}}, \bibinfo
  {author} {\bibfnamefont {R.}~\bibnamefont {Valette}},\ and\ \bibinfo {author}
  {\bibfnamefont {Y.}~\bibnamefont {Mesri}},\ }\href@noop {} {\bibfield
  {journal} {\bibinfo  {journal} {Computer Methods in Applied Mechanics and
  Engineering}\ }\textbf {\bibinfo {volume} {308}},\ \bibinfo {pages} {238}
  (\bibinfo {year} {2016})}\BibitemShut {NoStop}%
\bibitem [{\citenamefont {Luu}\ and\ \citenamefont {Forterre}(2009)}]{Luu-09}%
  \BibitemOpen
  \bibfield  {author} {\bibinfo {author} {\bibfnamefont {L.-H.}\ \bibnamefont
  {Luu}}\ and\ \bibinfo {author} {\bibfnamefont {Y.}~\bibnamefont {Forterre}},\
  }\href@noop {} {\bibfield  {journal} {\bibinfo  {journal} {Journal of Fluid
  Mechanics}\ ,\ \bibinfo {pages} {301}} (\bibinfo {year} {2009})}\BibitemShut
  {NoStop}%
\bibitem [{\citenamefont {Manglik}\ \emph {et~al.}(2001)\citenamefont
  {Manglik}, \citenamefont {Wasekar},\ and\ \citenamefont
  {Zhang}}]{Manglik-01}%
  \BibitemOpen
  \bibfield  {author} {\bibinfo {author} {\bibfnamefont {R.~M.}\ \bibnamefont
  {Manglik}}, \bibinfo {author} {\bibfnamefont {V.~M.}\ \bibnamefont
  {Wasekar}},\ and\ \bibinfo {author} {\bibfnamefont {J.}~\bibnamefont
  {Zhang}},\ }\href@noop {} {\bibfield  {journal} {\bibinfo  {journal}
  {Experimental thermal and fluid science}\ }\textbf {\bibinfo {volume} {25}},\
  \bibinfo {pages} {55} (\bibinfo {year} {2001})}\BibitemShut {NoStop}%
\bibitem [{\citenamefont {Boujlel}\ and\ \citenamefont
  {Coussot}(2013)}]{Boujlel-13}%
  \BibitemOpen
  \bibfield  {author} {\bibinfo {author} {\bibfnamefont {J.}~\bibnamefont
  {Boujlel}}\ and\ \bibinfo {author} {\bibfnamefont {P.}~\bibnamefont
  {Coussot}},\ }\href@noop {} {\bibfield  {journal} {\bibinfo  {journal} {Soft
  Matter}\ }\textbf {\bibinfo {volume} {9}},\ \bibinfo {pages} {5898} (\bibinfo
  {year} {2013})}\BibitemShut {NoStop}%
\bibitem [{\citenamefont {G\'eraud}\ \emph {et~al.}(2014)\citenamefont
  {G\'eraud}, \citenamefont {J{\o}rgensen}, \citenamefont {Petit},
  \citenamefont {Delano\"e-Ayari}, \citenamefont {Jop},\ and\ \citenamefont
  {Barentin}}]{Geraud-14}%
  \BibitemOpen
  \bibfield  {author} {\bibinfo {author} {\bibfnamefont {B.}~\bibnamefont
  {G\'eraud}}, \bibinfo {author} {\bibfnamefont {L.}~\bibnamefont
  {J{\o}rgensen}}, \bibinfo {author} {\bibfnamefont {L.}~\bibnamefont {Petit}},
  \bibinfo {author} {\bibfnamefont {H.}~\bibnamefont {Delano\"e-Ayari}},
  \bibinfo {author} {\bibfnamefont {P.}~\bibnamefont {Jop}},\ and\ \bibinfo
  {author} {\bibfnamefont {C.}~\bibnamefont {Barentin}},\ }\href@noop {}
  {\bibfield  {journal} {\bibinfo  {journal} {Europhysics Letters}\ }\textbf
  {\bibinfo {volume} {107}},\ \bibinfo {pages} {58002} (\bibinfo {year}
  {2014})}\BibitemShut {NoStop}%
\bibitem [{\citenamefont {J{\o}rgensen}\ \emph {et~al.}(2015)\citenamefont
  {J{\o}rgensen}, \citenamefont {{L}e {M}errer}, \citenamefont
  {Delano\"e-Ayari},\ and\ \citenamefont {Barentin}}]{Jorgensen-15}%
  \BibitemOpen
  \bibfield  {author} {\bibinfo {author} {\bibfnamefont {L.}~\bibnamefont
  {J{\o}rgensen}}, \bibinfo {author} {\bibfnamefont {M.}~\bibnamefont {{L}e
  {M}errer}}, \bibinfo {author} {\bibfnamefont {H.}~\bibnamefont
  {Delano\"e-Ayari}},\ and\ \bibinfo {author} {\bibfnamefont {C.}~\bibnamefont
  {Barentin}},\ }\href@noop {} {\bibfield  {journal} {\bibinfo  {journal} {Soft
  matter}\ }\textbf {\bibinfo {volume} {11}},\ \bibinfo {pages} {5111}
  (\bibinfo {year} {2015})}\BibitemShut {NoStop}%
\end{thebibliography}%


%apsrev4-2.bst 2019-01-14 (MD) hand-edited version of apsrev4-1.bst
%Control: key (0)
%Control: author (72) initials jnrlst
%Control: editor formatted (1) identically to author
%Control: production of article title (-1) disabled
%Control: page (0) single
%Control: year (1) truncated
%Control: production of eprint (0) enabled
%
%%%%%%%%%%%%%%%%%%%%%%%%%%%%%%%%%%%%%%%%%%%%%%%%%%%%%%%%%%%%%%%%%%%%%%%%%%%%%%
%%%%%%%%%%%%%%%%%%%%%%%%%%%%%%%%%%%%%%%%%%%%%%%%%%%%%%%%%%%%%%%%%%%%%%%%%%%%%%

\end{document}